\documentclass[aps,prd,nofootinbib]{revtex4}
\usepackage[colorlinks=true, pdfstartview=FitV, linkcolor=blue, citecolor=red, urlcolor=magenta]{hyperref}
\usepackage{graphicx}
\usepackage{latexsym}
\usepackage{amsmath}
\usepackage{amsfonts}
\usepackage{amssymb}
\usepackage{verbatim}
\usepackage{braket}
\usepackage{extarrows}
\usepackage{hyperref}

\begin{document}

\title{Casimir effect, loop corrections and topological mass generation for interacting real
and complex scalar fields in Minkowski spacetime with different conditions}
\author{$^{1}$A. J. D. Farias Junior}
\email{antonio.farias@academico.ufpb.br}
\author{$^{1}$Herondy F. Santana Mota}
\email{hmota@fisica.ufpb.br}
\affiliation{$^{1}$Departamento de F\' isica, Universidade Federal da Para\' iba,\\
Caixa Postal 5008, Jo\~ ao Pessoa, Para\' iba, Brazil.}

\begin{abstract}
In this paper the Casimir energy density, loop corrections, and generation of topological mass
are investigated for a system consisting of two interacting real and complex scalar fields. 
The interaction considered is the quartic interaction
in the form of a product of the modulus square of the complex field
and the square of the real field. In addition, it is also considered the 
self-interaction associated with each field. In this theory, the scalar field
is constrained to always obey periodic condition while the complex field
obeys in one case a quasiperiodic condition and in other case mixed boundary conditions.
The Casimir energy density, loop corrections,
and topological mass are evaluated analytically for the massive and massless scalar fields considered. An analysis of
possible different stable vacuum states and the corresponding stability 
condition is also provided. In order to better understand our investigation, some graphs
are also presented. The formalism we use here to perform such investigation is the effective
potential, which is written as loop expansions via path integral in quantum
field theory.

\end{abstract}

\maketitle

\section{Introduction}

Since its prediction in 1948, the Casimir effect is considered one of the
most interesting physical phenomenon. This effect, which is of pure quantum
nature, was predicted by H. Casimir \cite{casimir1948attraction}. In its
standard form, the Casimir effect consists in a force of attraction that
arises in a system of two neutral parallel and perfectly conducting plates,
placed in a classical vacuum, near to each other. This force of attraction is described
in the framework of quantum field theory and is due to modifications in the
vacuum fluctuations associated with the quantized electromagnetic field, as a consequence of the imposition 
of Dirichlet boundary condition on the plates. The phenomenon of the Casimir effect, in the case of the electromagnetic field,
has been confirmed by several high accuracy experiments \cite%
{bressi2002measurement, kim2008anomalies,
lamoreaux1997demonstration,lamoreaux1998erratum,mohideen1998precision,mostepanenko2000new,wei2010results}. 
Currently, it is also known that not only the electromagnetic field presents the Casimir
effect, but other fields as well, such as scalar and fermion fields. Additionally, the Casimir effect may also arise from
different boundary conditions. For instance, the Casimir effect associated
with a real scalar field subjected to a helix boundary condition with
temperature corrections is considered in Ref. \cite{aleixo2021thermal},
subjected to Robin boundary conditions in Ref. \cite{romeo2002casimir}, and
in Ref. \cite{maluf2020casimir} the Casimir energy for a real scalar field
and the Elko neutral spinor field in a field theory at a Lifshitz fixed
point is obtained. Moreover, in Ref. \cite{Haridev:2021jwi} a complex scalar field theory
has been considered and the corresponding Casimir energy density in compact
spacetimes investigated. A review on the Casimir effect can be found in Ref. \cite%
{bordag2009advances} (see also \cite%
{milton2001casimir,mostepanenko1997casimir}).

As we have mentioned, boundary conditions play a crucial role in the
investigation of the Casimir effect. A very interesting condition
is the quasiperiodic one. In Ref. \cite{feng2014casimir} a scalar field under a
quasiperiodic condition, inspired by nanotubes, is considered in
order to investigate the corresponding Casimir effect. It is found that the
Casimir force can be attractive or repulsive depending on the value of the
phase related to the quasiperiodic condition. Another
interesting boundary condition that modifies the quantum vacuum fluctuations
of a field is the one known as mixed boundary condition (Dirichlet -
Neumann). The Casimir energy arising as a consequence of the imposition of
mixed boundary conditions on a real self-interacting scalar field has been
considered in Ref. \cite{barone2004radiative} and in Ref. \cite{cruz2020casimir},
where a Lorentz violation scenario is also taken into account. Also, in Ref. \cite{FariasJunior:2022qsp}
it is studied a scalar field with a quartic self-interaction restricted to
obey a helix boundary condition.

The investigation of interacting quantum fields is important since
fields found in nature are always interacting. The interaction
mostly considered in the previous mentioned works is a quartic
self-interaction. In the framework of two interacting quantum fields, using the
effective potential approach, Toms in Ref. \cite{toms1980interacting} have considered two
real scalar fields, one twisted and the other untwisted, interacting via the
so called quartic interaction, i.e., the product between the square of the
two fields in a Euclidean spacetime, in order to investigate the
symmetry breaking and mass generation as a consequence of the nontrivial
topology produced by the periodic and antiperiodic conditions used. 
The quartic interaction between two real fields is also considered
in Ref. \cite{wang2022particle} for the study of the phenomenon of particle
production from oscillating scalar backgrounds in a Friedmann-Lema\^{\i}%
tre-Robertson-Walker universe using non-equilibrium quantum field theory. In this
type of calculation the renormalization procedure is of particular importance. In this
sense, Ref. \cite{wysozka2021two} presents a detailed discussion about the renormalization
scheme present in quantum field theory comprising two interacting scalar fields.

In order to investigate the Casimir energy density, loop corrections and generation of topological mass, 
in the present paper, we consider a system consisting of two real and complex
scalar fields interacting with each other via quartic interaction, in addition to the self-interactions which are normally present. The
real field is always subject to a periodic condition while the complex field is restricted 
to obey a quasiperiodic condition, as well as mixed boundary conditions used on
two identical and perfectly reflecting parallel planes
separated by a distance $L$. Hence,
we shall analyze separately two scenarios, one where a periodic real scalar field interacts 
with a complex scalar field obeying a quasiperiodic condition and the other
where a periodic real scalar field interacts 
with a complex scalar field obeying mixed boundary conditions. These two
scenarios generalize previous results found in the literature where it has been
considered self-interacting real scalar fields under periodic \cite{toms1980symmetry, Ford:1978ku} and
quasiperiodic \cite{porfirio2021ground} conditions, and also under mixed boundary conditions \cite{barone2004radiative, cruz2020casimir}. 
In this regard, we extend the analysis performed in Ref. \cite{toms1980interacting} to the complex scalar field and considering 
other conditions.

The choice of boundary conditions have to be mathematically
consistent and the choice of mixed boundary conditions are natural, for
instance, in the case of quantum gravity, spinor field theory and
supergravity \cite{luckock1991mixed}. Furthermore, the quasiperiodic
condition plays an important role when one considers nanotubes or
nanoloops for a quantum field \cite{feng2014casimir}. For example, if the
phase angle is zero (the periodic case), we have a corresponding system describing metallic nanotubes, while the values
$\pm \frac{2\pi }{3}$ for the phase angle correspond to semiconductor nanotubes. In our investigation,
we shall use the path integral formalism and construct the effective potential in
terms of loop expansions. This formalism was developed by Jackiw \cite%
{jackiw1974functional} and allows us to obtain the Casimir energy density
and loop corrections. The formalism also allows us, in principle, to calculate loop
corrections to the mass of the fields as a consequence of the nontrivial topology of the spacetime.
In our case, we consider only one-loop correction to the mass, which is enough to see
generation of topological mass at first order.

This paper is organized as follows: in Sec.\ref{sec2} we review the main
aspects of the path integral formalism to obtain the effective potential in
the case of two interacting quantum fields, one real and the other complex. 
The interaction considered is the quartic interaction, that
is, a product between the modulus square of the complex field and the square
of the real field. In addition, we also consider self-interaction
contributions for each field. In Sec.\ref{sec3}, we consider the real and complex
fields interacting with each other, where the real scalar field is subjected to a periodic
condition, while the components of the complex field obey a
quasiperiodic condition. By using the Riemann zeta function technique,
we evaluate the effective potential of the system, the Casimir
energy density and the one-loop correction to the mass. It is
also discussed the conditions for the stability of the vacuum states, which leads
to conditions for a positive topological mass. In Sec.\ref{sec3.2} the
components of the complex field will now be subjected to mixed boundary
conditions. The Casimir energy density, topological mass and the vacuum stability
are also investigated. In Sec.\ref{sec4} we present our conclusions. Through
this paper we use natural units in which both the Planck constant and the
speed of light are given by $\hslash =c=1$.

\section{Effective potential for interacting real and complex scalar fields}
\label{sec2}In this section we consider a real scalar field, $\psi $,
interacting with a complex scalar one, that is, $\phi$. The interaction term is in the
form of a product of the modulus square of the complex field and the square
of $\psi$. This choice of interaction satisfies
the discrete symmetry $\psi \rightarrow -\psi $, as well as the global symmetry $\phi \rightarrow e^{i\alpha
}\phi $. The existence of symmetries in particle physics is crucial for the predicability of a given model and
for a better fitting with
experimental data \cite{Haber:2019iyb}, making it important to consider this type of interactions in the investigation of the
Casimir effect, for instance. 
Note that it is also considered the
quartic self-interaction for each field. In the path integral approach for
the evaluation of the effective potential, it is usual to work with the
Euclidean spacetime coordinates, with imaginary time \cite%
{greiner2013field}. Moreover, the complex field $\phi $ can be decomposed in terms of
its real components as%
\begin{equation}
\phi =\frac{1}{\sqrt{2}}\left( \varphi _{1}+i\varphi _{2}\right) ,\qquad\qquad \phi
^{\ast }=\frac{1}{\sqrt{2}}\left( \varphi _{1}-i\varphi _{2}\right) .  \notag
\end{equation}%
The model associated with the system described above, in Euclidean coordinates, is
given by the following action:%
\begin{eqnarray}
&&S_{E}\left[ \psi ,\varphi _{i}\right] =\frac{1}{2}\int d^{4}x\left[ \psi
\square \psi +\sum_{i=1}^{2}\varphi _{i}\square \varphi _{i}\right] -\int
d^{4}xU\left( \psi ,\varphi \right) ,  \notag \\
&&U\left( \psi ,\varphi \right) =\frac{m^{2}+C_{2}}{2}\psi ^{2}+\frac{\mu
^{2}}{2}\varphi ^{2}+\frac{g}{2}\varphi ^{2}\psi ^{2}+\frac{\lambda
_{\varphi }}{4!}\varphi ^{4}+\frac{\lambda _{\psi }+C_{1}}{4!}\psi
^{4}+C_{3},  \label{rc2}
\end{eqnarray}%
where $m$ is the mass of the real field $\psi $, $\mu $ is the
mass of the complex field $\phi $, $\lambda _{\psi }$ and $\lambda _{\phi }$
are the coupling constants of self-interaction for the real and complex\ fields,
respectively, and $g$ is the coupling constant of the interaction between
the fields. The parameters $C_{i}$'s are the renormalization constants and their explicit
form will be obtained in the renormalization process of the effective
potential for each case considered in the next sections. We also make use of the notation $\varphi ^{2}=\varphi
_{1}^{2}+\varphi _{2}^{2}$. Furthermore, the d'Alembertian operator, $\square 
$, is written in Euclidean spacetime coordinates as%
\begin{equation}
\square =\left( \partial _{\tau }^{2}+\mathbf{\nabla }^{2}\right),
\label{dAl}
\end{equation}%
where $\tau=it$ is the imaginary time.

The construction of the effective potential using the path integral approach
is described in detail in Refs.~\cite{ryder1996quantum,toms1980symmetry}
(see also \cite{cruz2020casimir,porfirio2021ground}). Here we present only
the main steps necessary to our purposes. Thus, the action in Eq. (\ref{rc2}) is now
expanded about a fixed background $\Psi $ and $\Phi _{i}$ that is, $\psi =\Psi
+\chi $, $\varphi _{i}=\Phi _{i}+\varrho $, with $\chi $ and $\varrho $
representing quantum fluctuations. Since we are interested in the real field
 we seek to obtain the effective potential as a function only of $\Psi $,
i.e., $V_{\mathrm{eff}}\left( \Psi \right) $. Then it is unnecessary to
shift the components of the complex field $\phi $, which amounts to set $\Phi
_{i}=0$ \cite{jackiw1974functional}. Hence, we do not need to include
counter terms proportional to powers of $\varphi $ in Eq. \eqref{rc2}. A note here is in
place, if we try to carry out the shift in $\phi $, i.e., setting $\Phi
_{i}\neq 0$, a cross term will appear in the exponential on the r.h.s. of
Eq.~(\ref{rc2.3}), which turns the calculations needed for the analysis
extremely cumbersome. For this reason the simplification $\Phi _{i}=0$, is
justified in the calculations. In the next sections we will in fact impose 
on the components of the complex field a quasiperiodic condition as well as mixed boundary conditions.
This requires that we set the value $\Phi _{i}=0$, the only possible choice for a constant field to be
compatible with the imposed conditions. In these cases, the use of Eq. (\ref{rc2.3}) without cross terms 
becomes more accurate. Therefore, in the next sections, we shall be interested in analyzing the
influence of the complex scalar field, subjected to a quasiperiodic and mixed conditions, on both the Casimir energy
density and topological mass arising due to the real scalar field subjected to a periodic condition.

The expansion of the effective potential in powers of $\hslash $, up to
order $\hslash ^{2}$, can be written as%
\begin{equation}
V_{\mathrm{eff}}\left( \Psi \right) =V^{\left( 0\right) }\left( \Psi \right)
+V^{\left( 1\right) }\left( \Psi \right) +V^{\left( 2\right) }\left( \Psi
\right) .  \label{rc2.1}
\end{equation}%
The zero order term, $V^{\left( 0\right) }\left( \Psi \right) $, describes
the classical potential, i.e., the tree-level contribution,%
\begin{equation}
V^{\left( 0\right) }\left( \Psi \right) =U\left( \Psi \right) =\frac{%
m^{2}+C_{2}}{2}\Psi ^{2}+\frac{\lambda _{\psi }+C_{1}}{4!}\Psi ^{4}+C_{3}.
\label{rc2.2}
\end{equation}%
The next term, $V^{\left( 1\right) }\left( \Psi \right) $, is the one-loop
correction to the classical potential and, in terms of the path integral
approach, takes the following form \cite{toms1980interacting}:%
\begin{equation}
V^{\left( 1\right) }\left( \Psi \right) =-\frac{1}{\Omega _{4}}\ln \int 
\mathcal{D}\psi \mathcal{D}\varphi _{1}\mathcal{D}\varphi _{2}\exp \left\{ -%
\frac{1}{2}\left( \psi ,\hat{A}\psi \right) -\frac{1}{2}\left( \varphi _{1},%
\hat{B}\varphi _{1}\right) -\frac{1}{2}\left( \varphi _{2},\hat{B}\varphi
_{2}\right) \right\} ,  \label{rc2.3}
\end{equation}%
where $\Omega _{4}$ is the 4-dimensional volume of the Euclidean spacetime,
which depends on the conditions imposed on the fields. Note that
we have introduced the notation,%
\begin{equation}
\left( \psi ,\hat{A}\psi \right) =\int d^{4}x\psi \left( x\right) \hat{A}%
\psi \left( x\right) ,  \label{rc2.4}
\end{equation}%
with the self-adjoint operators $\hat{A}$ and $\hat{B}$ defined as%
\begin{equation}
\hat{A}=-\square +m^{2}+\frac{\lambda _{\psi }}{2}\Psi ^{2},\qquad\qquad \hat{B}%
=-\square +\mu ^{2}+g\Psi ^{2}.  \label{rc2.5}
\end{equation}%
The one-loop correction to the effective potential can be written in terms
of the eigenvalues of the operators $\hat{A}$ and $\hat{B}$, using the
generalized zeta function \cite{toms1980symmetry, aleixo2021thermal}. Let us denote by $\alpha
_{n}$ and $\beta _{n}$, the eigenvalues of the operators $\hat{A}$ and $\hat{%
B}$, respectively. Then, one can construct the generalized zeta function $%
\zeta \left( s\right) $ as follows%
\begin{equation}
\zeta _{\alpha }\left( s\right) =\sum_{\sigma}\alpha _{\sigma}^{-s},\qquad\qquad \zeta _{\beta
}\left( s\right) =\sum_{\rho}\beta _{\rho}^{-s},  \label{rc2.6}
\end{equation}%
where $\sigma$ and $\rho$ stand for the set of quantum numbers associated with the
eigenfunctions of the operators $\hat{A}$ and $\hat{B}$, respectively. The
summation symbol denotes sum or integration of the quantum numbers,
depending on whether they are discrete or continuous. It is possible to show that
the one-loop correction, (\ref{rc2.3}), can be written in terms of the
generalized zeta functions, (\ref{rc2.6}), as \cite%
{toms1980symmetry,Hhawking1977zeta}

\begin{eqnarray}
&&V^{\left( 1\right) }\left( \Psi \right) =V_{\alpha }^{\left( 1\right)
}\left( \Psi \right) +V_{\beta }^{\left( 1\right) }\left( \Psi \right) , 
\label{rc140}
\end{eqnarray}
where
\begin{eqnarray}
V_{\alpha }^{\left( 1\right) }=-\frac{1}{2\Omega _{4}}\left[ \zeta
_{\alpha }^{\prime }\left( 0\right) +\zeta _{\alpha }\left( 0\right) \ln \nu
^{2}\right] ,\qquad\qquad V_{\beta }^{\left( 1\right) }=-\frac{1}{\Omega _{4}}\left[
\zeta _{\beta }^{\prime }\left( 0\right) +\zeta _{\beta }\left( 0\right) \ln
\nu ^{2}\right] .  \label{rc14}
\end{eqnarray}%
In the above expressions, $\zeta _{\alpha ,\beta }\left( 0\right) $ and $\zeta
_{\alpha ,\beta }^{\prime }\left( 0\right) $ denote the generalized zeta
function and its derivative with respect to $s$, evaluated at $s=0$, respectively. Note that the parameter $\nu $ stands for an integration
measure in the functional space and is to be
removed via renormalization of the effective potential \cite{toms1980symmetry}. In addition, for practical
reasons, the two-loop correction, $V^{\left( 2\right) }\left( \Psi \right) $%
, of the effective potential is calculated from the two-loop graphs. This
correction can also be written in terms of the generalized zeta function
if one is interested in calculating the vacuum contribution \cite%
{cruz2020casimir,porfirio2021ground, FariasJunior:2022qsp}. We postpone the explicit form of $%
V^{\left( 2\right) }\left( \Psi \right) $ for later on, when we investigate it.

After one obtains the explicit form of the effective potential with its
corrections, it is required to renormalize it. The renormalization process
is achieved by means of a set of renormalization conditions. The first one
 is written in analogy to Coleman-Weinberg. It allows us to fix the
constant $C_{1}$ in Eq.~(\ref{rc2}) and also the coupling constant $\lambda
_{\psi }$, \cite{coleman1973radiative}. This condition is expressed as%
\begin{equation}
\left. \frac{d^{4}V_{\mathrm{eff}}\left( \Psi \right) }{d\Psi ^{4}}%
\right\vert _{\Psi =M}=\lambda _{\psi },  \label{rc14.2}
\end{equation}%
where $M$ is a parameter with dimension of mass, which in the case the model
is massive, we can take it as being zero \cite%
{cruz2020casimir,porfirio2021ground, FariasJunior:2022qsp}. 
The next renormalization condition which fix the constant $C_{2}$ in Eq.~(%
\ref{rc2}), is written as follows%
\begin{equation}
\left. \frac{d^{2}V_{\mathrm{eff}}\left( \Psi \right) }{d\Psi ^{2}}%
\right\vert _{\Psi =v}=m^{2},  \label{rc14.3}
\end{equation}%
where $v$ is the value that minimizes the effective potential.
It is pertinent to point out that the above expression also provides the
topological mass when we use the renormalized effective potential instead
of $V_{\mathrm{eff}}\left( \Psi \right) $. Note that $\Psi =v$ in Eq.~(%
\ref{rc14.3}) is the value of the field that minimizes the potential as long as the extremum
condition is obeyed
\begin{equation}
\left. \frac{dV_{\mathrm{eff}}\left( \Psi \right) }{d\Psi }\right\vert
_{\Psi =v}=0.  \label{rc14.4}
\end{equation}%
In Sec.\ref{sec3.1} we discuss the vacuum stability and present the
values of the field which satisfy the condition above. The last condition
one should use to renormalize the effective potential, fixing the constant $%
C_{3}$, is written in the form \cite{cruz2020casimir}%
\begin{equation}
\left. V_{\mathrm{eff}}\left( \Psi \right) \right\vert _{\Psi =0}=0,
\label{rc14.5}
\end{equation}%
which is relevant only if the model is massive \cite%
{cruz2020casimir,porfirio2021ground, FariasJunior:2022qsp}. It should be clear that the conditions presented in Eqs.~(\ref{rc14.2}), (%
\ref{rc14.3}) and (\ref{rc14.5}) are taken in the limit of Minkowski
spacetime.

We are now ready to study the loop expansion of the effective potential of
the real scalar field and the generation of topological mass, imposing a periodic condition for the real field, and a quasiperiodic
condition for the components of the complex field, along with mixed boundary
conditions. We shall consider the components of the complex field, $%
\varphi _{1}$ and $\varphi _{2}$, obeying the same boundary conditions. 

\section{Periodic and quasiperiodic conditions}
%
\label{sec3}We are considering a real field $\psi $,
interacting with a complex field $\phi $ via quartic interaction. The
action of the system is presented in Eq.~(\ref{rc2}). Note that the system
takes into consideration the quartic self-interaction terms as well. In this section, the
conditions which the fields must obey are the periodic, for the real field, and
quasiperiodic for the components of the complex field.

\begin{figure}[h]
\includegraphics[scale=0.25]{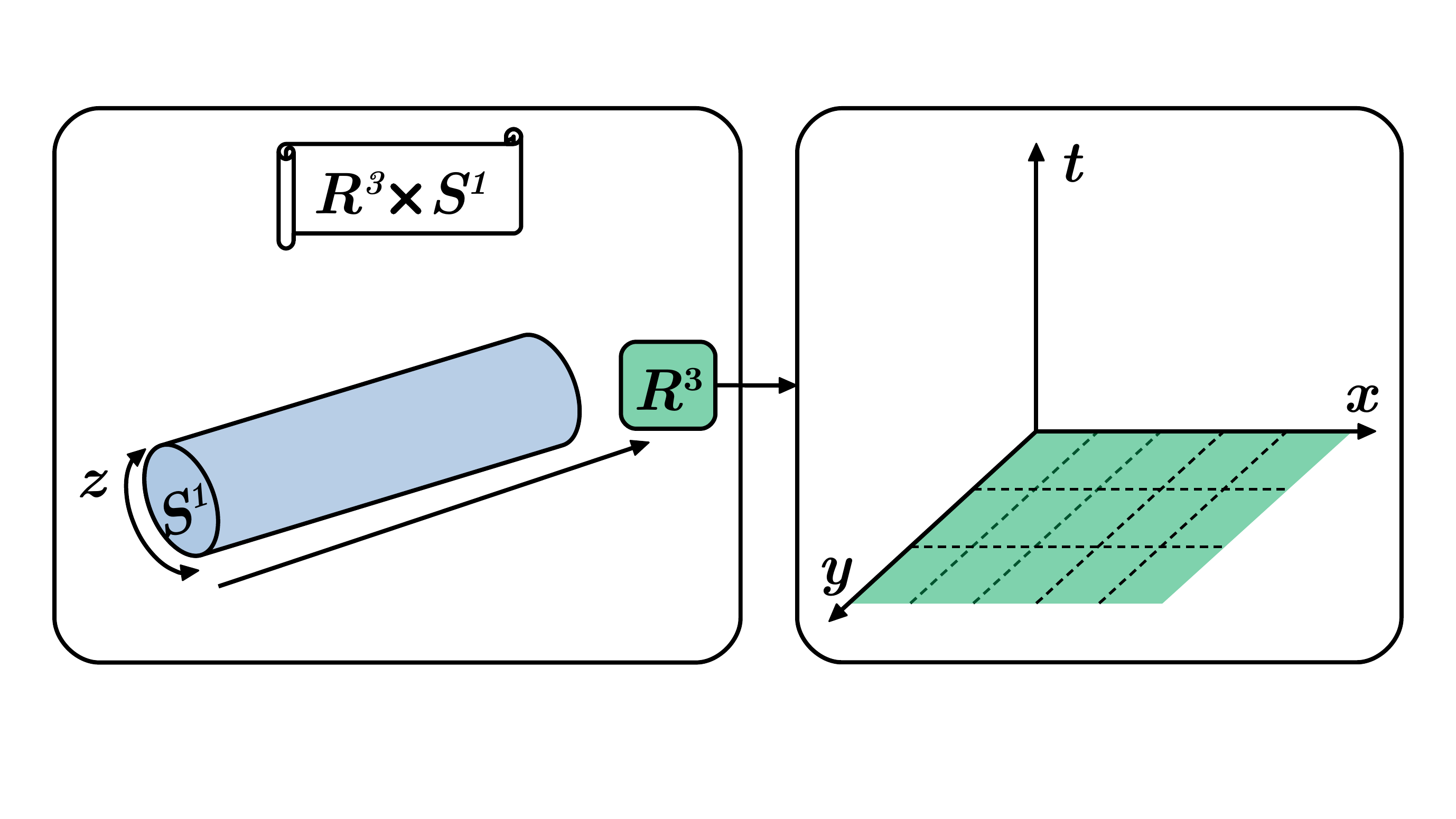}
\caption{Illustrative representation of a four-dimensional spacetime with a compactified spatial dimension. The spacetime is composed by the compactified spatial dimension $z$, $S^1$, and a tridimensional space $R^3$ of coordinates $t,x,y$.
}\label{figure1}
\end{figure}

The real field being subjected to the periodic condition means it must 
satisfy the following relation:%
\begin{equation}
\psi \left( \tau ,x,y,z+L\right) =\psi \left( \tau ,x,y,z\right) ,
\label{rc14.7}
\end{equation}%
where $L$ is the periodic parameter. In fact, the condition above leads to the compactification
of the $z$-coordinate into a length $L$, as show the illustration in Fig.\ref{figure1}. Hence, the
eigenvalues equation of the operator $\hat{A}$, presented in Eq.~(\ref{rc2.5}), is well known in the literature and
is written as%
\begin{equation}
\alpha _{\sigma }=k_{\tau }^{2}+k_{x}^{2}+k_{y}^{2}+ \frac{4\pi^2}{L^2} n^2+M_{\lambda }^{2},\qquad\qquad M_{\lambda }^{2}=m^{2}+\frac{\lambda
_{\psi }}{2}\Psi ^{2}, \label{rc14.1}
\end{equation}%
where $n=0,\pm 1,\pm 2,..., $ and the subscript $\sigma $ stands for the set of quantum numbers $\left(
k_{\tau },k_{x},k_{y},n\right) $.
For the components of the complex field, $\varphi _{i}$, we apply the
quasiperiodic condition \cite{feng2014casimir, porfirio2021ground}, i.e.,
\begin{equation}
\varphi _{i}\left( \tau ,x,y,z+L\right) =e^{2i\pi \theta }\varphi _{i}\left(
\tau ,x,y,z\right) .  \label{rc14.8}
\end{equation}%
This condition also compactifies the $z$-coordinate into a length $L$, but now there exists the
influence of the phase $\theta$, that is, the quasiperiodic parameter that assumes values in the
range $0\le\theta<1$. In this sense, the quasiperiodic condition recovers the periodic one for $\theta=0$.
The case for $\theta=1/2$ recovers the well known antiperiodic condition. Thereby, under the 
quasiperiodic condition, the eigenvalues of the
operator $\hat{B}$, presented in Eq.~(\ref{rc2.5}), take the form%
\begin{equation}
\beta _{\rho }=p_{\tau }^{2}+p_{x}^{2}+p_{y}^{2}+\frac{4\pi^2}{L^{2}}\left( n+\theta \right) ^{2}+M_{g}^{2},\qquad\qquad M_{g}^{2}=\mu
^{2}+g\Psi ^{2},  \label{rc16.1}
\end{equation}%
where $n=0,\pm 1,\pm 2,..., $ and $\rho $ stands for the set of
the quantum numbers $\left( p_{\tau },p_{x},p_{y},n\right) $. The values $\theta =0$
and $\theta =1/2$ are also known as the cases for the untwisted and twisted
scalar fields, respectively.

Knowing the explicit form of the eigenvalues $\alpha _{\sigma }$ and $\beta
_{\rho }$, given in Eqs. (\ref{rc14.1}) and (\ref{rc16.1}), respectively,
one can construct the generalized zeta function from Eq.~(\ref{rc2.6}) and
obtain a practical expression for the first order correction to the effective
potential in Eq.~(\ref{rc14}). We shall do that next, also obtaining 
the topological mass.

\subsection{One-loop correction}
Starting from the eigenvalues presented in Eq.~(\ref{rc14.1}), which are
associated with the real scalar field $\psi $, we construct the generalized zeta
function from Eq.~(\ref{rc2.6}) as%
\begin{equation}
\zeta _{\alpha }\left( s\right) =\frac{\Omega _{3}}{\left( 2\pi \right) ^{3}}%
\int dk_{\tau }dk_{x}dk_{y}\sum_{n=-\infty }^{+\infty }\left\{ k_{\tau
}^{2}+k_{x}^{2}+k_{y}^{2}+\left( \frac{2\pi n}{L}\right) ^{2}+M_{\lambda
}^{2}\right\} ^{-s},  \label{rc1.1}
\end{equation}%
where $\Omega _{3}$ stands for the 3-dimensional volume associated with the
Euclidean spacetime coordinates $\tau ,x,y$, necessary to make the integrals
dimensionless. In order to obtain an expression for the generalized
zeta function (\ref{rc1.1}), we shall follow similar steps as the ones presented in 
\cite{porfirio2021ground, FariasJunior:2022qsp}. We keep most of the calculation for the
convenience of the reader. Thus, by using the identity,%
\begin{equation}
w^{-s}=\frac{2}{\Gamma \left( s\right) }\int_{0}^{\infty }d\tau \ \tau
^{2s-1}e^{-w\tau ^{2}},  \label{i1}
\end{equation}%
in Eq. \eqref{rc1.1} and performing the resulting Gaussian integrals in $k_{\tau }$, $k_{x}$ and 
$k_{z}$, one obtains the generalized zeta function in the form%
\begin{equation}
\zeta _{\alpha }\left( s\right) =\frac{\Omega _{3}}{\left( 2\pi \right) ^{2}}%
\frac{\pi ^{\frac{1}{2}}}{\Gamma \left( s\right) }\sum_{n=-\infty }^{+\infty
}\int_{0}^{\infty }d\tau \ \tau ^{2s-4}\exp \left\{ -\tau ^{2}\left[ \left( 
\frac{2\pi n}{L}\right) ^{2}+M_{\lambda }^{2}\right] \right\} .  \label{z}
\end{equation}%
The expression obtained in Eq.~(\ref{z}) is suited for the use of the well
known integral representation of the gamma function $\Gamma \left( z\right) $
\cite{abramowitz1965handbook}%
\begin{equation}
\Gamma \left( z\right) =2\int_{0}^{\infty }d\mu \ \mu ^{2z-1}e^{-\mu ^{2}},
\label{intg}
\end{equation}%
which allows us to rewrite the generalized zeta function (\ref{z}) in terms
only of the summation in $n$, i.e.,%
\begin{equation}
\zeta _{\alpha }\left( s\right) =\frac{\Omega _{4}\pi ^{\frac{3}{2}-2s}}{%
2^{2s}L^{4-2s}}\frac{\Gamma \left( s-\frac{3}{2}\right) }{\Gamma \left(
s\right) }\sum_{n=-\infty }^{+\infty }\left[ n^{2}+\left( \frac{M_{\lambda }L%
}{2\pi }\right) ^{2}\right] ^{\frac{3}{2}-s}.  \label{z2}
\end{equation}%
The quantity $\Omega _{4}$ stands for the 4-dimensional volume in Euclidean
spacetime, which takes into account the spacetime topology $S^1\times R^3$ 
as a consequence of the periodic condition imposed on the field. In the case
under consideration, the 4-dimensional volume is written as $\Omega
_{4}=\Omega _{3}L$. In order to perform the sum in Eq.~(\ref{z2}) we use
the following analytic continuation of the inhomogeneous generalized Epstein
function \cite{elizalde1995zeta,feng2014casimir, FariasJunior:2022qsp}:%
\begin{equation}
\sum_{n=-\infty }^{+\infty }\left[ \left( n+\vartheta \right) ^{2}+\kappa
^{2}\right] ^{-z}=\frac{\pi ^{\frac{1}{2}}\kappa ^{1-2z}}{\Gamma \left(
z\right) }\left\{ \Gamma \left( z-\frac{1}{2}\right) +4(\pi \kappa )^{z-%
\frac{1}{2}}\sum_{j=1}^{\infty }j^{z-\frac{1}{2}}\cos \left( 2\pi j\vartheta
\right) K_{\left( \frac{1}{2}-z\right) }\left( 2\pi j\kappa \right) \right\}
,  \label{id2}
\end{equation}%
where $K_{\gamma }(x)$ is the modified Bessel function of the second kind
or, as it is also known, the Macdonald function \cite{abramowitz1965handbook}. 
After the use of Eq.~(\ref{id2}), the generalized zeta function in
Eq.~(\ref{z2}) is presented in the form%
\begin{equation}
\zeta _{\alpha }\left( s\right) =\frac{\Omega _{4}M_{\lambda }^{4-2s}}{%
2^{4}\pi ^{2}\Gamma \left( s\right) }\left\{ \Gamma \left( s-2\right)
+2^{4-s}\sum_{j=1}^{\infty }f_{\left( 2-s\right) }\left( jM_{\lambda
}L\right) \right\} ,  \label{rc1.2}
\end{equation}%
where we have defined the function $f_{\mu }\left( x\right) $ as%
\begin{equation}
f_{\gamma }\left( x\right) =\frac{K_{\gamma }\left( x\right) }{x^{\gamma }}.
\label{rc1.4}
\end{equation}%

By evaluating the generalized zeta function in Eq.$~$(\ref{rc1.2}) and its
derivative with respect to $s$, in the limit $s\rightarrow 0$, one finds from Eq.~(\ref{rc14})
the one-loop contribution to the effective
potential as
\begin{equation}
V_{\alpha }^{\left( 1\right) }\left( \Psi \right) =\frac{M_{\lambda }^{4}}{%
64\pi ^{2}}\left[ \ln \left( \frac{M_{\lambda }^{2}}{\nu ^{2}}\right) -\frac{%
3}{2}\right] -\frac{M_{\lambda }^{4}}{2\pi ^{2}}\sum_{j=1}^{\infty
}f_{2}\left( jM_{\lambda }L\right) .  \label{rc1.3}
\end{equation}%
The expression above is the first order correction, that is, the one-loop
correction to the effective potential due to the periodic condition. It remains to be
evaluated the contribution due to the complex field for the one-loop correction, which we shall analyze below.

For the components of the complex scalar field, the eigenvalues are presented in
Eq.~(\ref{rc16.1}), allowing us to construct the associated generalized zeta function in the form%
\begin{equation}
\zeta _{\beta }\left( s\right) =\frac{\Omega _{3}}{\left( 2\pi \right) ^{3}}%
\int dp_{\tau }dp_{x}dp_{y}\sum_{n=-\infty }^{+\infty }\left[ p_{\tau
}^{2}+p_{x}^{2}+p_{y}^{2}+\frac{4\pi^2}{L^{2}}\left(
n+\theta \right) ^{2}+M_{g}^{2}\right] ^{-s}.  \label{rc1.5}
\end{equation}%
The expression for the generalzized zeta function, $\zeta _{\beta }\left( s\right) $, 
arising due to a real scalar field has been obtained in
detail in \cite{porfirio2021ground} and, of course, for our 
case is very similar since the componentes of the complex 
field considered are real. Therefore, we present only the final result, i.e.,%
\begin{equation}
\zeta _{\beta }\left( s\right) =\frac{\Omega _{4}M_{g}^{4-2s}}{16\pi
^{2}\Gamma \left( s\right) }\left\{ \Gamma \left( s-2\right)
+2^{4-s}\sum_{j=1}^{\infty }\cos \left( 2\pi j\theta \right) f_{\left(
2-s\right) }\left( jM_{g}L\right) \right\} .  \label{rc1.6}
\end{equation}%
From Eq.~(\ref{rc14}) and using the generalized zeta function presented above, one is able to write
the first order correction to the effective potential due to the complex field as%
\begin{equation}
V_{\beta }^{\left( 1\right) }\left( \Psi \right) =\frac{M_{g}^{4}}{32\pi ^{2}%
}\left[ \ln \left( \frac{M_{g}^{2}}{\nu ^{2}}\right) -\frac{3}{2}\right] -%
\frac{M_{g}^{4}}{\pi ^{2}}\sum_{j=1}^{\infty }\cos \left( 2\pi j\theta
\right) f_{2}\left( jM_{g}L\right) .  \label{rc1.7}
\end{equation}

Collecting the results presented in Eqs.~(\ref{rc1.3}) and (\ref{rc1.7}),
we can write the first order correction to the effective
potential associated with the interacting real and complex scalar fields in the following form:%
\begin{eqnarray}
V^{\left( 1\right) }\left( \Psi \right) &=&\frac{M_{\lambda }^{4}}{64\pi ^{2}%
}\left[ \ln \left( \frac{M_{\lambda }^{2}}{\nu ^{2}}\right) -\frac{3}{2}%
\right] +\frac{M_{g}^{4}}{32\pi ^{2}}\left[ \ln \left( \frac{M_{g}^{2}}{\nu
^{2}}\right) -\frac{3}{2}\right] +  \notag \\
&&-\frac{M_{\lambda }^{4}}{2\pi ^{2}}\sum_{j=1}^{\infty }f_{2}\left(
jM_{\lambda }L\right) -\frac{M_{g}^{4}}{\pi ^{2}}\sum_{j=1}^{\infty }\cos
\left( 2\pi j\theta \right) f_{2}\left( jM_{g}L\right) .  \label{rc1.8}
\end{eqnarray}%
Therefore, from Eq.~(\ref{rc140}), the nonrenormalized effective potential up to one-loop correction reads%
\begin{eqnarray}
V_{\mathrm{eff}}\left( \Psi \right) &=&\frac{m^{2}+C_{2}}{2}\Psi ^{2}+\frac{%
\lambda _{\psi }+C_{1}}{4!}\Psi ^{4}+C_{3}+  \notag \\
&&+\frac{M_{\lambda }^{4}}{64\pi ^{2}}\left[ \ln\left( \frac{M_{\lambda }^{2}}{\nu
^{2}}\right) -\frac{3}{2}\right] +\frac{M_{g}^{4}}{32\pi ^{2}}\left[ \ln \left( 
\frac{M_{g}^{2}}{\nu ^{2}}\right) -\frac{3}{2}\right] +  \notag \\
&&-\frac{M_{\lambda }^{4}}{2\pi ^{2}}\sum_{j=1}^{\infty }f_{2}\left(
jM_{\lambda }L\right) -\frac{M_{g}^{4}}{\pi ^{2}}\sum_{j=1}^{\infty }\cos
\left( 2\pi j\theta \right) f_{2}\left( jM_{g}L\right) .  \label{rcq2}
\end{eqnarray}%
It is clear that the one-loop corrections in Eqs. (\ref{rc1.3}) and (\ref%
{rc1.7}) for the cases of real and complex fields, respectively, differ by a 
factor of two when $\theta=0$, the periodic condition particular case. This is justified since the complex
field has two components. Note that the masses are also, in general, different. That is, for the real
scalar field is $m$ and for the complex scalar field is $\mu$.

Once we obtain the effective potential $V_{\mathrm{eff}}\left( \Psi \right) $
in Eq.~(\ref{rcq2}), our task is now to renormalize it. Hence, by following the
renormalization procedure, from the conditions presented in Eqs.~(\ref%
{rc14.2}), (\ref{rc14.3}) and (\ref{rc14.5}), in the limit of Minkowski
spacetime $L\rightarrow \infty $, one obtains the renormalization constants
as%
\begin{eqnarray}
&&C_{1}=\frac{3\lambda _{\psi }^{2}}{32\pi ^{2}}\ln \left( \frac{\nu ^{2}}{%
m^{2}}\right) +\frac{3g^{2}}{4\pi ^{2}}\ln \left( \frac{\nu ^{2}}{m^{2}}%
\right) ,  \notag \\
&&C_{2}=\frac{\lambda _{\psi }m^{2}}{32\pi ^{2}}\left[ \ln \left( \frac{\nu
^{2}}{m^{2}}\right) +1\right] +\frac{g\mu ^{2}}{8\pi ^{2}}\left[ \ln \left( 
\frac{\nu ^{2}}{\mu ^{2}}\right) +1\right] ,  \notag \\
&&C_{3}=\frac{m^{4}}{64\pi ^{2}}\left[ \ln \frac{\nu ^{2}}{m^{2}}+\frac{3}{2}%
\right] +\frac{\mu ^{4}}{32\pi ^{2}}\left[ \ln \left( \frac{\nu ^{2}}{\mu
^{2}}\right) +\frac{3}{2}\right] .  \label{rcq3}
\end{eqnarray}%
Furthermore, by substituting the renormalization constants above into the
nonrenormalized effective potential given in Eq.~(\ref{rcq2}), we are able to write
the renormalized effective potential in the following form:%
\begin{eqnarray}
V_{\mathrm{eff}}^{R}\left( \Psi \right) &=&\frac{m^{2}}{2}\Psi ^{2}+\frac{%
\lambda _{\psi }}{4!}\Psi ^{4}+\frac{\mu ^{4}}{32\pi ^{2}}\ln \left( \frac{%
M_{g}^{2}}{\mu ^{2}}\right) +\frac{m^{4}}{64\pi ^{2}}\ln \left( \frac{%
M_{\lambda }^{2}}{m^{2}}\right) +  \notag \\
&&+\frac{g\mu ^{2}\Psi ^{2}}{16\pi ^{2}}\left[ \ln \left( \frac{M_{g}^{2}}{%
\mu ^{2}}\right) -\frac{1}{2}\right] +\frac{\lambda _{\psi }^{2}\Psi ^{4}}{%
256\pi ^{2}}\left[ \ln \left( \frac{M_{\lambda }^{2}}{m^{2}}\right) -\frac{3%
}{2}\right] +  \notag \\
&&+\frac{\lambda _{\psi }m^{2}\Psi ^{2}}{64\pi ^{2}}\left[ \ln \left( \frac{%
M_{\lambda }^{2}}{m^{2}}\right) -\frac{1}{2}\right] +\frac{g^{2}\Psi ^{4}}{%
32\pi ^{2}}\left[ \ln \left( \frac{M_{g}^{2}}{\mu ^{2}}\right) -\frac{3}{2}%
\right] +  \notag \\
&&-\frac{M_{\lambda }^{4}}{2\pi ^{2}}\sum_{j=1}^{\infty }f_{2}\left(
jM_{\lambda }L\right) -\frac{M_{g}^{4}}{\pi ^{2}}\sum_{j=1}^{\infty }\cos
\left( 2\pi j\theta \right) f_{2}\left( jM_{g}L\right) .  \label{rcq4}
\end{eqnarray}%
The explicit form of the renormalized effective potential presented in Eq.~(%
\ref{rcq4}) makes possible to evaluate the Casimir energy density and also
the topological mass, up to first order correction.

\begin{figure}[h]
\includegraphics[scale=0.3]{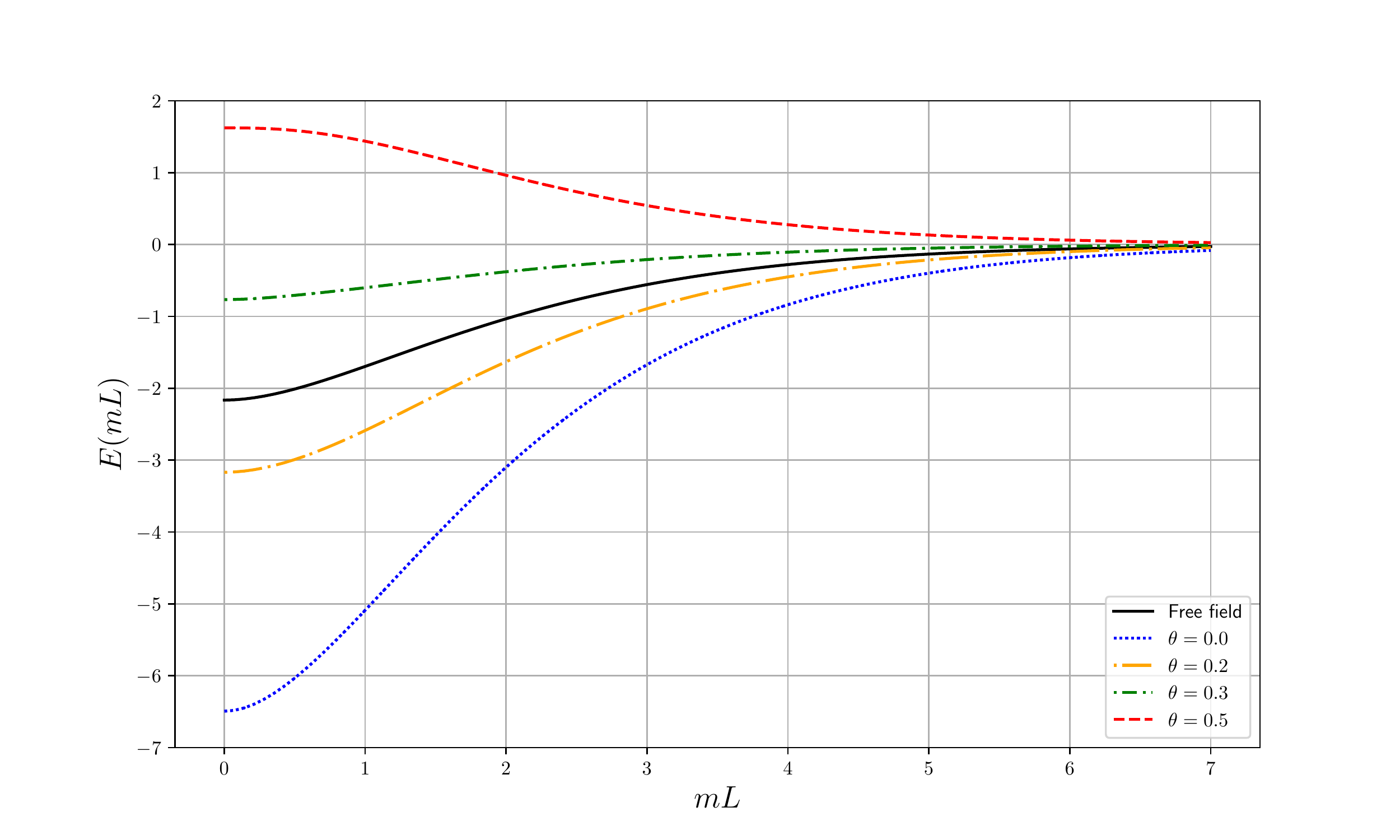}
\includegraphics[scale=0.3]{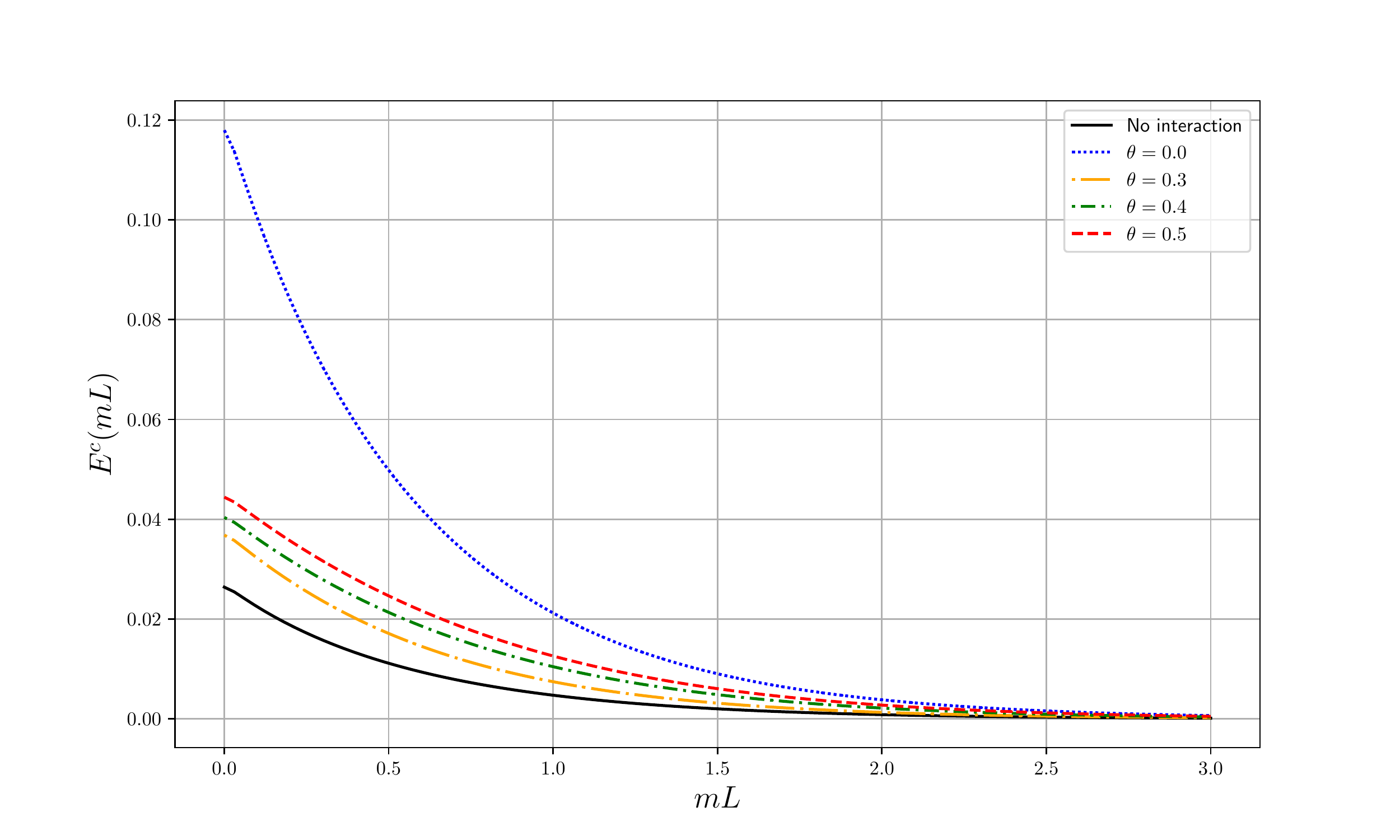}
\caption{Plot of the dimensionless Casimir energy density, $E(mL)=2\pi^2L^4\mathcal{E}_{\mathrm{C}}$, defined from Eq. (\ref{rc1.15}), as a function of $mL$, is shown on the left. The plot on the right shows the dimensionless two-loop contribution to the Casimir energy density, $E^c(mL)=32\pi^4L^4\Delta\mathcal{E}_{\mathrm{C}}$, defined from Eq. \eqref{c2l4}, as a function of $mL$ and considering $\lambda_{\psi}=10^{-2}$,
$\lambda_{\varphi}=10^{-2}$ and $g=10^{-3}$. For both cases we have taken $\mu=m$ and different values of $\theta$.}\label{figure2}
\end{figure}

In order to proceed and calculate the vacuum energy density, let us
consider $\Psi=0$ as the stable vacuum state of the theory, although
there are other possible stable vacuum states, as analyzed in Sec.\ref{sec3.1}. 
Thus, from the renormalized effective potential $V_{\mathrm{eff}}^{R}\left( \Psi
\right) $ in Eq. (\ref{rcq4}) one can evaluate the Casimir energy density in a
straightforwardly way since the vacuum state is obtained by setting $\Psi =0$.
Hence, the Casimir energy density is found to be%
\begin{eqnarray}
\mathcal{E}_{\mathrm{C}}&=&\left. V_{\mathrm{eff}}^{R}\left( \Psi \right)
\right\vert _{\Psi =0}\nonumber\\
&=&-\frac{m^{4}}{2\pi ^{2}}\sum_{n=1}^{\infty
}f_{2}\left( nmL\right) -\frac{\mu ^{4}}{\pi ^{2}}\sum_{j=1}^{\infty }\cos
\left( 2\pi j\theta \right) f_{2}\left( j\mu L\right) .  \label{rc1.15}
\end{eqnarray}%
Note that the first term on the r.h.s. of Eq.~(\ref{rc1.15}) is the
contribution to the Casimir energy density from the real scalar field $\psi $ subjected to
a periodic condition, while the second term is the
contribution from the complex field subjected to a quasiperiodic condition \cite%
{porfirio2021ground}. In the particular case, $\theta=0$, this contribution is twice
the one from the real scalar field if the masses are equal. In order to show the influence
of the complex field, under a quasiperiodic condition, on the Casimir energy density
of the real field, we have plotted the expression in Eq. \eqref{rc1.15} as a function of $mL$
which is shown on the left side of Fig.\ref{figure2} for different values of $\theta$ and taken $\mu=m$. The black solid line is the Casimir energy density
free of interaction with the complex field, only with the effect of the real field self-interaction. It is clear that
depending on the value of the quasiperiodic parameter $\theta$, the Casimir energy density
can be bigger or smaller than the free case, including the possibility 
of assuming positive or negative values. Note that the curves tend to 
repeat their behavior for values such that $\theta>0.5$. For instance, the curve 
represented by the green dot-dashed line for $\theta=0.3$ is the same as 
the one for $\theta=0.7$, and so on. Furthermore, all the curves end in their corresponding 
massless field constant value cases at $mL=0$, as it can be checked from Eq. \eqref{rc1.16}. Also,
in the regime $mL\gg 1$, the Casimir energy density in Eq. \eqref{rc1.15} goes to zero for all curves,
as revealed by the plot on the left side of Fig.\ref{figure2}. This a consequence of the exponentially suppressed
behavior of the Macdonald function for large arguments \cite%
{abramowitz1965handbook}. 

It is interesting to consider the case of
massless scalar fields, i.e., the limit $m,\mu \rightarrow 0$ of Eq. (\ref{rc1.15}). For the
massless scalar field case, we can make use of the limit for small arguments of the
Macdonald function, i.e., $K_{\mu }\left( x\right) \simeq \frac{\Gamma
\left( \mu \right) }{2}\left( \frac{2}{x}\right) ^{\mu }$ \cite%
{abramowitz1965handbook}. Hence, from Eq.~(\ref{rc1.15}), we obtain the
Casimir energy density for interacting massless scalar fields as%
\begin{equation}
\mathcal{E}_{\mathrm{C}}=-\frac{\pi ^{2}}{90L^{4}}-\frac{2}{\pi ^{2}L^{4}}%
\sum_{j=1}^{\infty }j^{-4}\cos \left( 2\pi j\theta \right) ,
\label{rc1.15.a}
\end{equation}%
where we have used the following result for the Riemann zeta function $\zeta \left(
4\right) =\frac{\pi ^{4}}{90}$ \cite{elizalde1995zeta,elizalde1995ten}, on
the first term on the r.h.s. of Eq.~(\ref{rc1.15.a}). This term is the
already known Casimir effect result for a free massless scalar field 
subjected to a periodic condition \cite{toms1980symmetry}. In addition, 
the second term on the r.h.s. of Eq.~(\ref{rc1.15.a}) can be
rewritten in terms of the well known Bernoulli polynomials,%
\begin{equation}
B_{2k}\left( \theta \right) =\frac{\left( -1\right) ^{k-1}2\left( 2k\right) !%
}{\left( 2\pi \right) ^{2k}}\sum_{n=1}^{\infty }\frac{\cos \left( 2\pi
n\theta \right) }{n^{2k}}.  \label{rc1.15.b}
\end{equation}%
Hence, the expression for the Casimir energy density,
in the massless scalar fields case, is found to be%
\begin{equation}
\mathcal{E}_{\mathrm{C}}=-\frac{\pi ^{2}}{90L^{4}}+2\frac{\pi ^{2}}{3L^{4}}%
\left( \theta ^{4}-2\theta ^{3}+\theta ^{2}-\frac{1}{30}\right) ,
\label{rc1.16}
\end{equation}%
where we have made use of the Bernoulli polynomial of fourth order, that is, $B_{4}\left(
\theta \right) =\left( \theta ^{4}-2\theta ^{3}+\theta ^{2}-\frac{1}{30}%
\right) $. As one should expect, the first term on the r.h.s. of Eq.~(\ref%
{rc1.16}) is consistent with the result found in \cite{toms1980symmetry}
while the second term is consistent with the result found in \cite%
{feng2014casimir, porfirio2021ground} (taking into account the two components of the complex
field). The latter also provides the right expressions for the
periodic $\left( \theta =0\right) $ and antiperiodic $\left( \theta =\frac{1}{2}%
\right) $ cases, also known as untwisted and twisted cases, respectively. 

We wish now to investigate the influence of the conditions in the
mass $m$ of the real scalar field, i.e., the generation of topological mass at
one-loop level. From the condition presented in Eq.~(\ref{rc14.3}) and making use
of the renormalized effective potential \eqref{rcq4}, one
obtains the following expression for the topological mass of the real scalar field $%
\psi $:
\begin{equation}
m_{T}^{2}=m^{2}\left[ 1+\frac{\lambda _{\psi }}{4\pi ^{2}}\sum_{n=1}^{\infty
}f_{1}\left( nmL\right) +\frac{\mu ^{2}}{m^{2}}\frac{g}{\pi ^{2}}%
\sum_{j=1}^{\infty }\cos \left( 2\pi j\theta \right) f_{1}\left( j\mu
L\right) \right] .  \label{rcq5}
\end{equation}%
Note that the topological mass $m_{T}^{2}$ does not present any divergencies, making possible for us to
consider the massless field limit, that is, $m,\mu \rightarrow 0$. Hence, by
using the same approximation for the Macdonald function as the one applied to obtain Eq.~(\ref{rc1.15.a}) from Eq. \eqref{rc1.15},
we find the topological mass as%
\begin{equation}
m_{T}^{2}=\frac{\lambda _{\psi }}{24L^{2}}+\frac{g}{L^{2}}\left( \theta
^{2}-\theta +\frac{1}{6}\right) .  \label{rcq6}
\end{equation}%
In the expression above we have used the Riemann zeta function $\zeta \left( 2\right) =\frac{%
\pi ^{2}}{6}$ \cite{elizalde1995zeta,elizalde1995ten} and also the Bernoulli
polynomials presented in Eq.~(\ref{rc1.15.b}), with $%
B_{2}\left( \theta \right) =\theta ^{2}-\theta +\frac{1}{6}$. As one can
see, the mass correction comes from the self-interaction term, which is
proportional to $\lambda _{\psi }$, and also from the interaction between
the fields, which is proportional to the coupling constant $g$. Note that the first term
on the r.h.s. of Eq. \eqref{rcq6} has been previously obtained in Ref. \cite{toms1980symmetry}
in a real scalar field theory with only self-interaction. 

Another interesting 
aspect associated with the topological mass is that if $\lambda _{\psi }<g$,
Eq. \eqref{rcq6} may become negative depending on the value of $\theta$,
which would in principle indicate vacuum instability. Had we, for instance, considered
a complex scalar field theory with only self-interaction (no interaction between the fields)
this would be a problem since it does not make sense to
consider a constant complex field, $\Phi_i\ne0$, compatible with the quasiperiodic condition
for $\theta\ne 0$. The case $ \theta= 0$ is not problematic in this regard and that is
why we have made the real scalar field to obey the periodic condition. Nevertheless, this problem is solved by taking into account an interaction theory as the one considered 
 here (see also \cite{toms1980interacting}). Within this theory it is possible to study the 
 vacuum stability, which in fact is made in Sec.\ref{sec3.1} by considering for simplicity a massless scalar field theory. The analysis indicates that
 the vacuum $\Psi=0$ is stable only if $\lambda_{\psi}>-24gB_2(\theta)$, otherwise 
 it is necessary to consider the two other possible stable vacuum states, $\Psi_{\pm}$, in Eq. \eqref{vs7}.
\begin{figure}[h]
\includegraphics[scale=0.3]{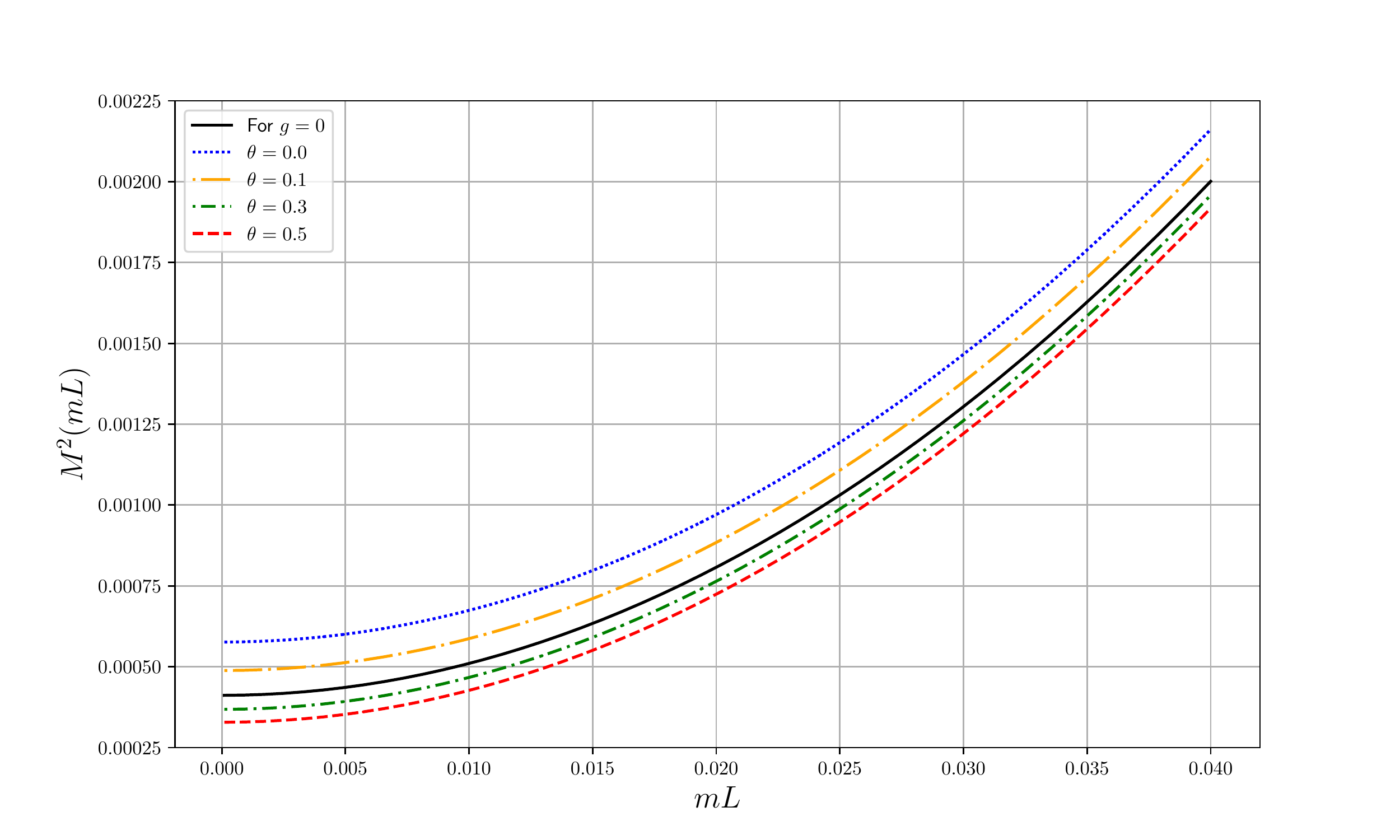}
\includegraphics[scale=0.3]{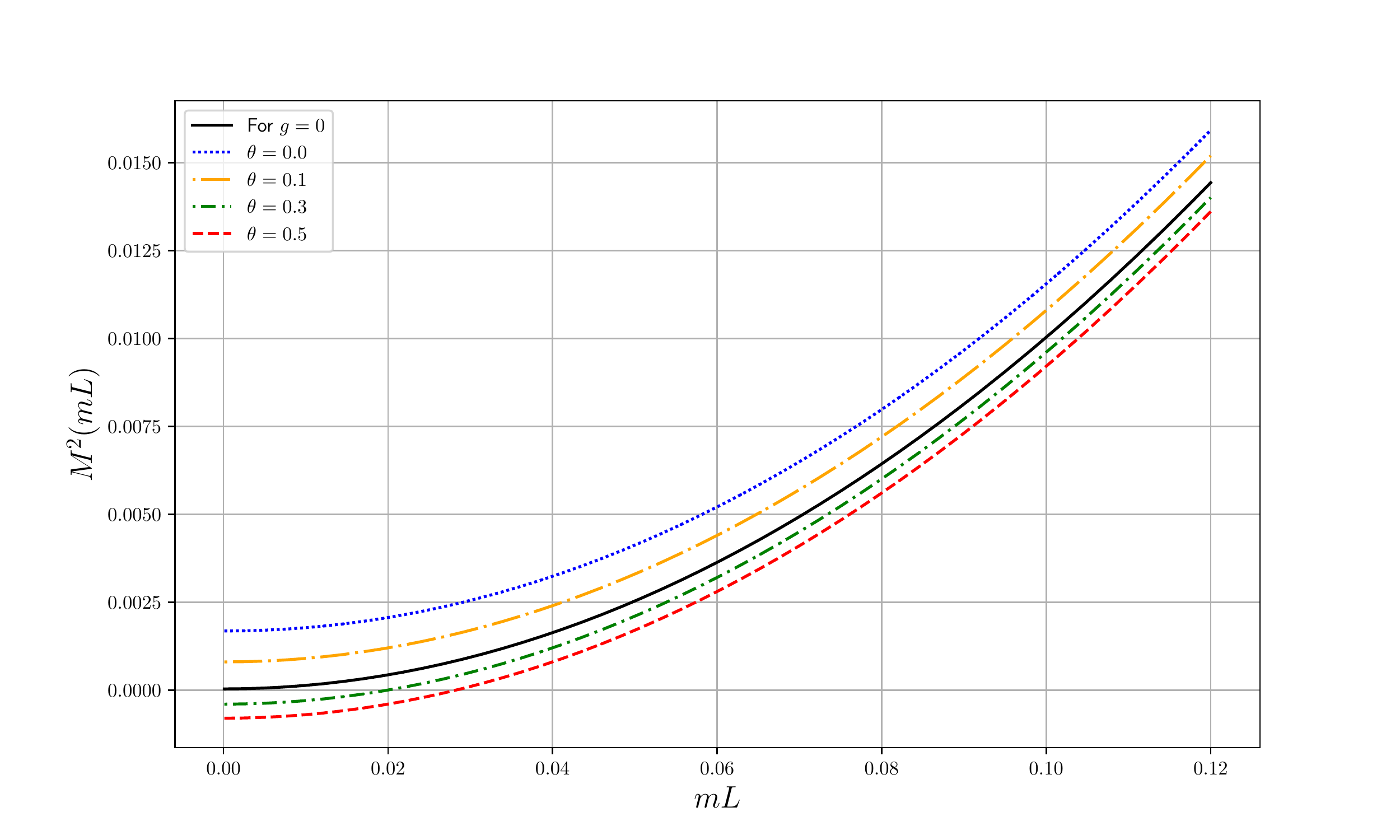}
\caption{Plot of the dimensionless topological mass squared, $M^2(mL)=m_T^2L^2$, defined from Eq. \eqref{rcq5}, as a function of $mL$, for different values of $\theta$ and taken $\mu=m$. On the left, the plot shows the curves for $\lambda_{\psi}= 10^{-2}$ and $g= 10^{-3}$ while on the right the plot shows the curves for $\lambda_{\psi}= 10^{-3}$ and $g= 10^{-2}$.}\label{figure3}
\end{figure}

In Fig.\ref{figure3} we have plotted the dimensionless mass squared, $M^2(mL)=m_T^2L^2$, 
defined from Eq. \eqref{rcq5}, as a function of $mL$, for different values of $\theta$ and taken $\mu=m$. On the 
left of Fig.\ref{figure3} the plot shows the curves for $\lambda_{\psi}= 10^{-2}$ and $g= 10^{-3}$, which satisfies 
the condition $\lambda_{\psi}>-24gB_2(\theta)$ in order $\Psi=0$ be a stable vacuum. This plot 
provides positive values in Eq. \eqref{rcq6} for all values of the quasiperiodic parameter $\theta$, as we should expect. In contrast, the plot
on the right shows the curves for $\lambda_{\psi}= 10^{-3}$ and $g= 10^{-2}$, which satisfies 
the condition $\lambda_{\psi}<-24gB_2(\theta)$. In this case, there exist negative values of Eq. \eqref{rcq6}
 for $\theta=0.3$ and $\theta=0.5$, showing that $\Psi=0$ is in fact an unstable vacuum state. However,
in a massive scalar field theory, $\Psi=0$, becomes stable for larger values of $mL$ even if $\lambda _{\psi }<g$, as indicates the 
plot on the right. Note
that all curves, at $mL=0$, end in their corresponding constant massless scalar field values for the topological mass in Eq. \eqref{rcq6}. 
For large values of $mL$ the Macdonald function is exponentially suppressed and the curves are dominated by
the first term on the r.h.s. of Eq. \eqref{rcq5}. Note also that the curves tend to repeat themselves for $\theta>0.5$.

The one-loop correction analysis is now done, so one can proceed to the
two-loop correction contribution by still considering $\Psi=0$ as the stable vacuum state.
As we now know, it means that we have to consider the restriction $\lambda_{\psi}>-24gB_2(\theta)$.

\subsection{Two-loop correction}
%
\label{sec3.3}We want now to analyze the loop correction to the Casimir energy density obtained in Eqs. \eqref{rc1.15}
and \eqref{rc1.15.a}. This can be done by considering the second order correction to the effective potential,
which can be obtained from the two-loop Feynman graphs. Since we have more than one contribution, we evaluate all the
two-loop contributions from each Feynman graph separately. Hence, we write
\begin{equation}
V^{\left( 2\right) }\left( \Psi \right) =V_{\lambda _{\psi }}^{\left(
2\right) }\left( \Psi \right) +V_{\lambda _{\varphi }}^{\left( 2\right)
}\left( \Psi \right) +V_{g}^{\left( 2\right) }\left( \Psi \right)
+V_{2\lambda _{\varphi }}^{\left( 2\right) }\left( \Psi \right),
\label{2lc}
\end{equation}%
where $V_{\lambda _{\psi }}^{\left(
2\right) }$ is the contribution from the self-interaction term, $\frac{\lambda _{\psi }}{4!}\psi ^{4}$, of the real
field, $V_{\lambda _{\varphi
}}^{\left( 2\right) }$ is the contribution from the self-interaction of the
complex field, that is, $\frac{\lambda _{\varphi }}{4!}\varphi _{1}^{4}$ and $\frac{%
\lambda _{\varphi }}{4!}\varphi _{2}^{4}$, $V_{g}^{\left( 2\right) }$ is
associated with the interaction between the real and complex fields $\frac{g}{%
2}\varphi _{1}^{2}\psi ^{2}$ and $\frac{g}{2}\varphi _{2}^{2}\psi ^{2}$ and,
finally, $V_{2\lambda _{\varphi }}^{\left( 2\right) }$ is associated with the
cross terms of the components of the complex field $\frac{\lambda _{\varphi }%
}{4!}2\varphi _{1}^{2}\varphi _{2}^{2}$.

\begin{figure}[h]
\includegraphics[scale=0.3]{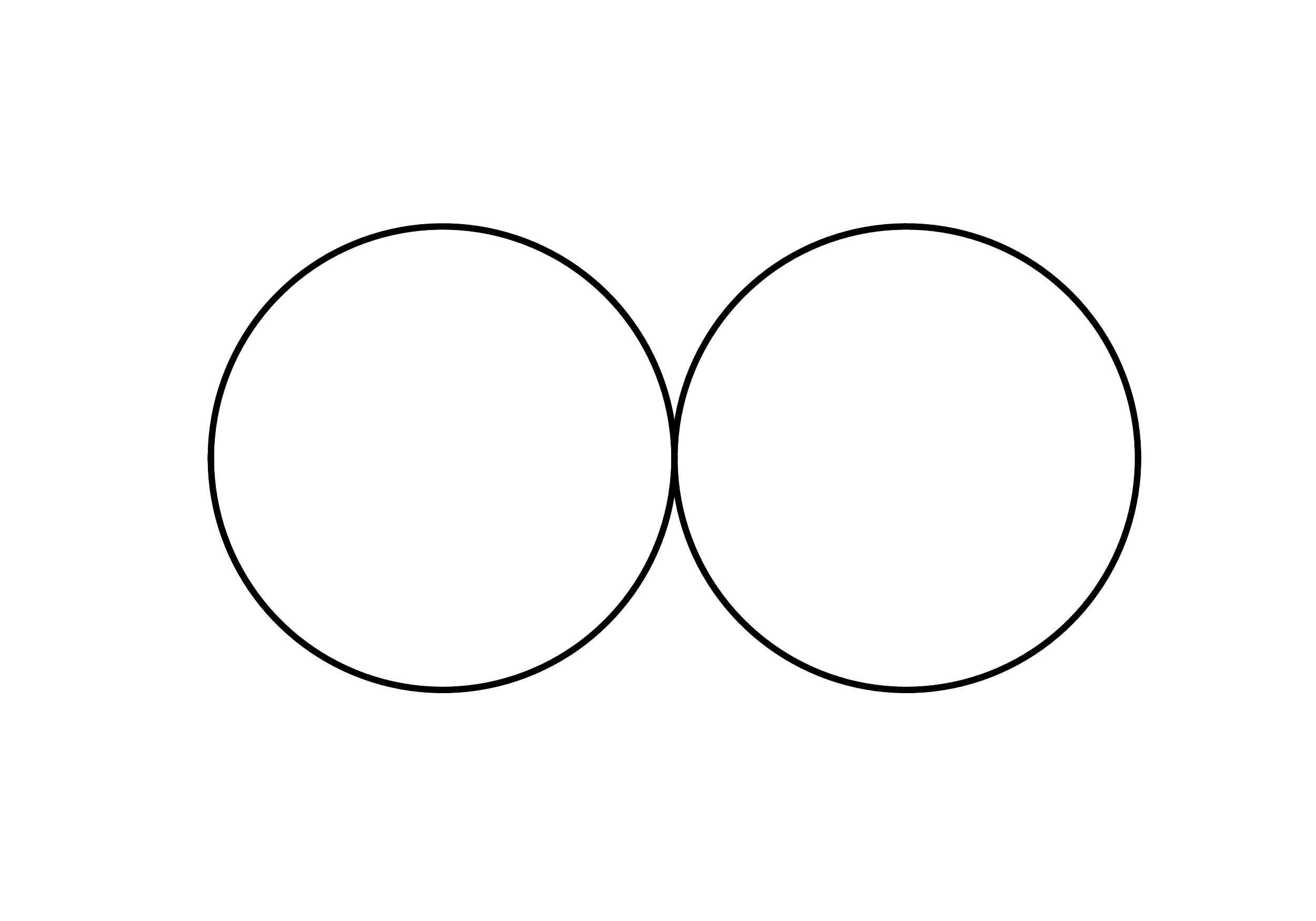}
\caption{Feynman graph representing the only non-vanishing self-interaction contribution to the two-loop correction calculated at $\Psi=0$.}\label{figure4}
\end{figure}

Let us first consider the contribution from the self-interaction term
associated with the real field, $V_{\lambda _{\psi }}^{\left( 2\right)
}\left( \Psi \right) $. Since we are interested in the vacuum state where $%
\Psi =0$, the only nonvanishg contribution comes from the graph exhibited
in Fig.\ref{figure4}. With the help of this Feynman graph one can write the two-loop
contribution in terms of the generalized zeta function presented in Eq.~(\ref%
{rc1.2}) in the following form \cite{FariasJunior:2022qsp, porfirio2021ground}:
\begin{equation}
V_{\lambda _{\psi }}^{\left( 2\right) }\left( 0\right) =\frac{\lambda _{\psi
}}{8}\left. \left[ \frac{\zeta _{\alpha }^{R}\left( 1\right) }{\Omega _{4}}%
\right] ^{2}\right\vert _{\Psi =0}.  \label{rcq7}
\end{equation}%
The zeta function $\zeta _{\alpha }^{R}\left( s\right) $ is defined as the
non-divergent part of the generalized zeta function given by Eq.~(\ref{rc1.2}%
), at $s=1$ \cite{FariasJunior:2022qsp,porfirio2021ground}, i.e.,%
\begin{equation}
\zeta _{\alpha }^{R}\left( s\right) =\zeta _{\alpha }\left( s\right) -\frac{%
\Omega _{4}M_{\lambda }^{4-2s}}{16\pi ^{2}}\frac{\Gamma \left( s-2\right) }{%
\Gamma \left( s\right) }.  \label{rcq8}
\end{equation}%
Note that the term that is being subtracted in the above equation is
independent of the parameter $L$, when divided by $\Omega _{4}$, characterizing the conditions and, as usual, should be dropped.
Explicitly, one obtains the following result for the two-loop contribution due to
the self-interaction term of the real field:%
\begin{equation}
V_{\lambda _{\psi }}^{\left( 2\right) }\left( 0\right) =\frac{\lambda _{\psi
}m^{4}}{32\pi ^{4}}\left[ \sum_{j=1}^{\infty }f_{1}\left( jmL\right) \right]
^{2}.  \label{rcq9}
\end{equation}%
The above result shows that the two-loop contribution is proportional to
the coupling constant $\lambda _{\psi }$, as it should.

The second contribution, $V_{\lambda _{\varphi}}^{\left( 2\right) }\left( \Psi \right) $, 
to the total two-loop correction in Eq. \eqref{2lc} can be read from the same graph
as the one in Fig.\ref{figure4}. We can construct the function $\zeta _{\beta }^{R}\left( s\right) 
$ from the generalized zeta function (\ref{rc1.6}), subtracting the
divergent part at $s=1$, and then obtain the contribution from the
self-interaction of the complex field, that is,%
\begin{equation}
V_{\lambda _{\varphi }}^{\left( 2\right) }\left( 0\right) =2\frac{\lambda
_{\varphi }\mu ^{4}}{32\pi ^{4}}\left[ \sum_{j=1}^{\infty }\cos \left( 2\pi
j\theta \right) f_{1}\left( j\mu L\right) \right] ^{2},  \label{rcq11}
\end{equation}%
which is proportional to $\lambda _{\varphi }$. The factor of two accounts for the two components of the
complex field, that is, $\varphi _{1}$ and $\varphi _{2}$ that give rise to
equal contributions.

\begin{figure}[h]
\includegraphics[scale=0.3]{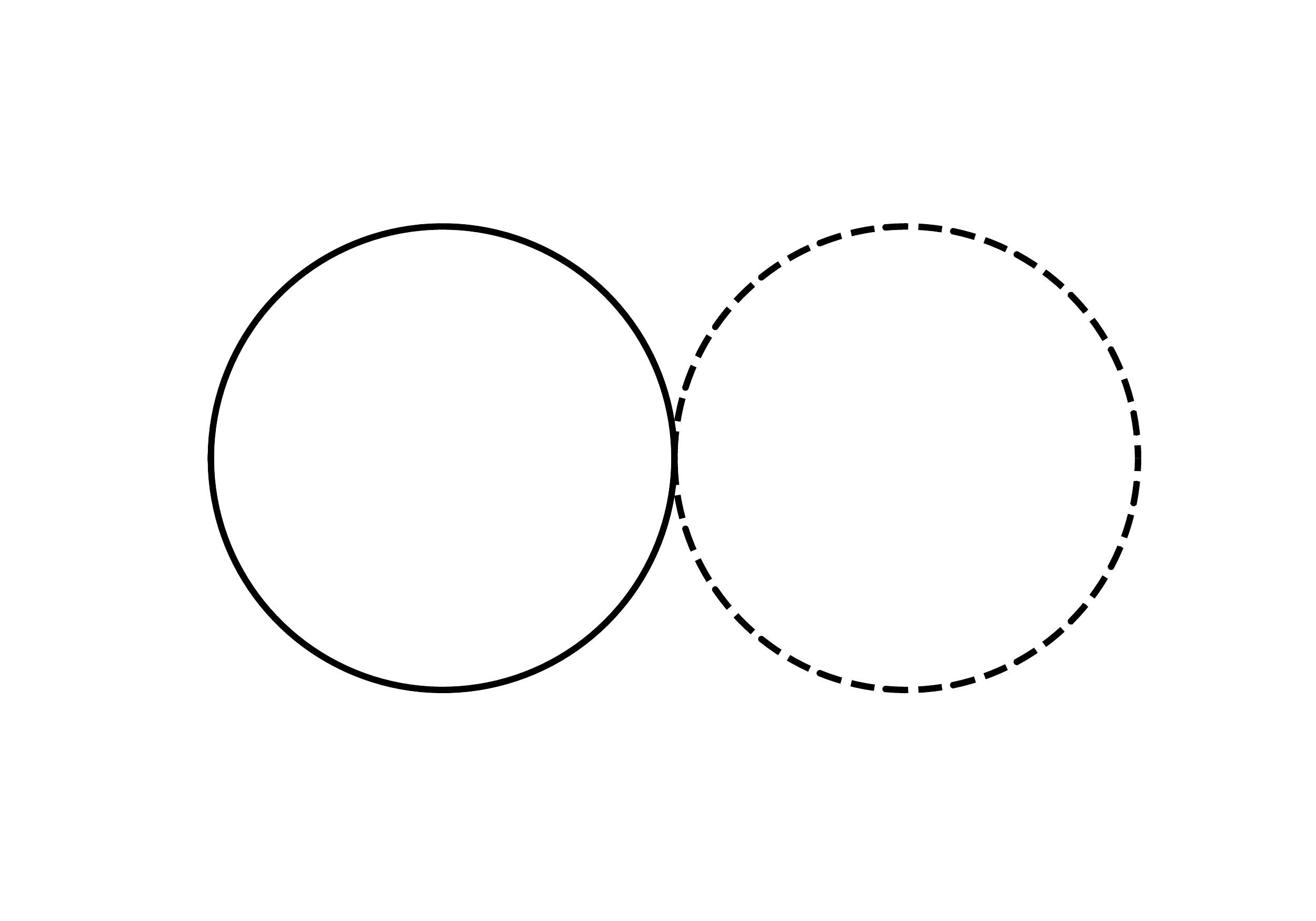}
\caption{Feynman graph representing the only non-vanishing contribution to the two-loop correction, calculated at $\Psi=0$, due to the interaction between the real and complex fields. This graph also provides the only non-vanishing contribution due to the interaction of the components of the complex field. In this case, the solid line represents the propagator associated with the component $\varphi _{1}$ while the dashed line represents the propagator 
associated with the component $\varphi _{2}$.}\label{figure5}
\end{figure}

Next we analyze the contributions from the interaction, 
$V_{g}^{\left( 2\right) }\left( \Psi \right) $, between the fields. This correction can be read
from the graph shown in Fig.\ref{figure5} and calculated, at $s=1$, by using the non-divergent part 
of the zeta functions in Eqs.~(\ref{rc1.2}) and (\ref{rc1.6}). Hence, one finds the contribution
from the interaction between the fields as%
\begin{equation}
V_{g}^{\left( 2\right) }\left( 0\right) =2\frac{gm^{2}\mu ^{2}}{8\pi ^{4}}%
\left[ \sum_{n=1}^{\infty }f_{1}\left( nmL\right) \right] \left[
\sum_{j=1}^{\infty }\cos \left( 2\pi j\theta \right) f_{1}\left( j\mu
L\right) \right] .  \label{v2fp}
\end{equation}%
Since the expression (\ref{v2fp}) comes from the interaction term, it is
proportional to the coupling constant $g$.

Finally, the last contribution, $V_{2\lambda _{\varphi }}^{\left( 2\right)
}\left( \Psi \right) $, comes from the interaction of the
components of the complex field (also a self-interaction). The contribution from this term is obtained from the
graph in Fig.\ref{figure5}, considering the solid line as representing the
propagator associated with the field $\varphi _{1}$ and the dashed one
associated with the field $\varphi _{2}$. Then, by using again the non-divergent
part of the zeta function (\ref{rc1.6}), calculated at $s=1$, the result is written as
\begin{equation}
V_{2\lambda _{\varphi }}^{\left( 2\right) }\left( 0\right) =\frac{\lambda
_{\varphi }\mu ^{4}}{48\pi ^{4}}\left[ \sum_{j=1}^{\infty }\cos \left( 2\pi
j\theta \right) f_{1}\left( j\mu L\right) \right] ^{2}.  \label{v2ff}
\end{equation}%
Note that, likewise the result presented in Eq.~(\ref{rcq11}), the above expression is proportional to $\lambda _{\varphi }$.

Collecting all the results obtained in Eqs~(\ref{rcq9}), (\ref{v2fp}), (\ref%
{rcq11}) and (\ref{v2ff}), one can write the total two-loop correction to the
effective potential in Eq. \eqref{2lc}, at the vacuum state $\Psi =0$, as%
\begin{eqnarray}
\Delta\mathcal{E}_{\mathrm{C}} &=&\left. V^{\left( 2\right) }\left( \Psi \right) \right\vert _{\Psi =0} \nonumber\\
&=&\frac{\lambda _{\psi }m^{4}}{32\pi ^{4}}\left[ \sum_{j=1}^{\infty
}f_{1}\left( jmL\right) \right] ^{2}+\frac{\lambda _{\varphi }\mu ^{4}}{%
12\pi ^{4}}\left[ \sum_{j=1}^{\infty }\cos \left( 2\pi j\theta \right)
f_{1}\left( j\mu L\right) \right] ^{2}+  \notag \\
&&+\frac{gm^{2}\mu ^{2}}{4\pi ^{4}}\left[ \sum_{n=1}^{\infty }f_{1}\left(
nmL\right) \right] \left[ \sum_{j=1}^{\infty }\cos \left( 2\pi j\theta
\right) f_{1}\left( j\mu L\right) \right] .  \label{c2l4}
\end{eqnarray}%
Therefore, combining the results presented in Eqs.~(\ref{rc1.15}) and (\ref%
{c2l4}), one obtains a correction to the Casimir energy density in Eq. \eqref{rc1.15}, 
which is first order in all coupling constants of the theory. As we can notice, while
the first order correction to the efective potential gives the Casimir energy density
associated with a free scalar and complex fields theory, the second order 
correction to the effective potential in Eq. \eqref{c2l4} provides a contribution to the 
Casimir energy density that is linearly proportional to all coupling constants.

Moreover, one can also consider the massless scalar fields limit of Eq. \eqref{c2l4}, that is,
$\mu,m\rightarrow 0$. This gives 
\begin{equation}
\Delta\mathcal{E}_{\mathrm{C}} = \frac{\lambda _{\psi }}{1152L^{4}}+\frac{\lambda _{\varphi }}{%
12L^{4}}\left( \theta ^{2}-\theta +\frac{1}{6}\right) ^{2}+\frac{g}{24L^{4}}%
\left( \theta ^{2}-\theta +\frac{1}{6}\right) .  \label{c4ims}
\end{equation}%
Hence, the expression above is the first order correction, in all coupling constants
of the theory, to the massless Casimir energy density in Eq. (\ref{rc1.16}). Note that
the first term on the r.h.s. of Eq. \eqref{c4ims} has been obtained in Ref. \cite{toms1980symmetry},
whilst the second term is consistent with the result obtained in Ref. \cite{porfirio2021ground} 
in a real scalar field theory considering only self-interaction. In contrast, the third term is a new one arising
from the interaction between the fields.

In Fig.\ref{figure2}, the plot on the right shows the influence of the complex field, under a quasiperiodic condition,
on the correction \eqref{c2l4} to the Casimir energy density. Thus, the expression in Eq. \eqref{c2l4} has been
plotted as a function of $mL$, for different values of $\theta$ and taken $\mu=m$. We also have considered $\lambda_{\psi}=10^{-2}$,
$\lambda_{\varphi}=10^{-2}$ and $g=10^{-3}$. The black solid line is the correction
free of interaction with the complex field, only with the effect of the real field self-interaction. It is clear that the curves
for $\theta\ne 0$ increase the correction when compared to the black solid line. Note that the curves here also tend to 
repeat their behavior for values such that $\theta>0.5$. For instance, the curve 
represented by the orange dot-dashed line for $\theta=0.3$ is the same as 
the one for $\theta=0.7$. Furthermore, all the curves tend to their corresponding 
massless field constant value cases at $mL=0$, as it can be checked from Eq. \eqref{c4ims}. Also,
in the regime $mL\gg 1$, the correction in Eq. \eqref{c2l4} goes to zero for all curves,
as revealed by the plot on the right side of Fig.\ref{figure2}. This a consequence of the exponentially suppressed
behavior of the Macdonald function for large arguments \cite%
{abramowitz1965handbook}.

Next, we shall analyze the vacuum stability of the theory, since the state $\Psi =0$ is not the
only possible vacuum state, as we have already anticipated.

%
\subsection{Vacuum stability}
%
\label{sec3.1}We want to analyze here the stability of the possible vacuum states associated with the effective potential, up to first order loop correction, of the theory described by the action in Eq. \eqref{rc2}.
For simplicity we consider the case where the fields are massless, i.e., $%
m,\mu \rightarrow 0$. It is import to point out again that, for the complex scalar field obeying the condition in Eq. (\ref%
{rc14.8}), the only constant field that can satisfy such a condition is the zero field, hence, we set $\Phi _{i}=0$.
This fact also turns the approximation discussed below Eq. \eqref{dAl} into an exact expression, namely, 
the one in Eq. (\ref{rc2.3}), which does not consider cross terms.

By following the same steps as the ones to obtain Eq. \eqref{rcq2}, the nonrenormalized effective potential for the massless scalar fields case is
written as%
\begin{eqnarray}
&&V_{\mathrm{eff}}\left( \Psi \right) =\frac{\lambda _{\psi }+C}{4!}\Psi
^{4}+\frac{\lambda _{\psi }^{2}\Psi ^{4}}{256\pi ^{2}}\left[ \ln\left( \frac{%
\lambda _{\psi }\Psi ^{2}}{2\nu ^{2}}\right)-\frac{3}{2}\right] +\frac{g^{2}\Psi
^{4}}{32\pi ^{2}}\left[ \ln \left( \frac{g\Psi ^{2}}{\nu ^{2}}\right) -\frac{%
3}{2}\right] +  \notag \\
&&-\frac{\lambda _{\psi }^{2}\Psi ^{4}}{8\pi ^{2}}\sum_{j=1}^{\infty
}f_{2}\left( j\sqrt{\frac{\lambda _{\psi }}{2}\Psi ^{2}}L\right) -\frac{%
g^{2}\Psi ^{4}}{\pi ^{2}}\sum_{j=1}^{\infty }\cos \left( 2\pi j\theta
\right) f_{2}\left( j\sqrt{g\Psi ^{2}}L\right) .  \label{vs1}
\end{eqnarray}%
Furthermore, the condition which takes care of the renormalization constant $C$, is given by Eq. \eqref{rc14.2}. 
Thereby, by applying this condition
 on the effective
potential of Eq.~(\ref{vs1}), in the Minkowski limit $L\rightarrow\infty$, one finds that the constant $C$ is given by%
\begin{equation}
C=\frac{3\lambda _{\psi }^{2}}{32\pi ^{2}}\ln\left( \frac{2\nu ^{2}}{\lambda
_{\psi }M ^{2}}\right)+\frac{3g^{2}}{4\pi ^{2}}\ln\left( \frac{\nu ^{2}}{gM ^{2}}\right)-%
\frac{\lambda _{\psi }^{2}}{4\pi ^{2}}-\frac{2g^{2}}{\pi ^{2}}.  \label{vs3}
\end{equation}%
Next, by substituting $C$ in the effective potential (\ref%
{vs1}), we obtain the renormalized effective potential for the massless scalar
fields theory, i.e.,%
\begin{eqnarray}
&&V_{\mathrm{eff}}^{R}\left( \Psi \right) =\frac{\lambda _{\psi }}{4!}\Psi
^{4}+\left[ \frac{\lambda _{\psi }^{2}}{8}+g^{2}\right] \frac{\Psi ^{4}}{%
32\pi ^{2}}\ln\left( \frac{\Psi ^{2}}{M ^{2}}\right)-\left[ \frac{\lambda _{\psi }^{2}}{%
8}+g^{2}\right] \frac{25\Psi ^{4}}{192\pi ^{2}}+  \notag \\
&&-\frac{\lambda _{\psi }^{2}\Psi ^{4}}{8\pi ^{2}}\sum_{j=1}^{\infty
}f_{2}\left( j\sqrt{\frac{\lambda _{\psi }}{2}\Psi ^{2}}L\right) -\frac{%
g^{2}\Psi ^{4}}{\pi ^{2}}\sum_{j=1}^{\infty }\cos \left( 2\pi j\theta
\right) f_{2}\left( j\sqrt{g\Psi ^{2}}L\right) .  \label{vs4}
\end{eqnarray}

Let us now investigate the possible vacuum states of the above renormalized effective potential, up to first order in the
coupling constants $\lambda _{\psi }$ and $g$, which is more than enough since we have considered corrections to the
Casimir energy density as well as to the mass of the scalar field only up to first order in the coupling constants. Thus, expanding the renormalized
effective potential given by Eq.~(\ref{vs4}) in powers of $\lambda _{\psi }$
and $g$ \cite{toms1980interacting}, up to first order, results in the
following expression:%
\begin{equation}
V_{\mathrm{eff}}^{R}\left( \Psi \right)\simeq-\frac{\pi ^{2}%
}{90L^{4}}+\frac{2\pi ^{2}}{3L^{4}}B_{4}\left( \theta \right) +\frac{\lambda
_{\psi }}{4!}\Psi ^{4}+\frac{\Psi ^{2}}{48L^{2}}\left[ \lambda _{\psi
}+24gB_{2}\left( \theta \right) \right] ,  \label{vs5}
\end{equation}%
where $B_{4}\left( \theta \right) $ and $B_{2}\left( \theta \right) $ are
the Bernoulli polynominals defined in Eq.~(\ref{rc1.15.b}). The minimum of
the potential, which corresponds to the vacuum state, is obtained as usual
by taking its derivative and equating the resulting expression to zero, that
is,%
\begin{equation}
\frac{\lambda _{\psi }}{6}\Psi ^{3}+\frac{\Psi }{24L^{2}}\left[ \lambda
_{\psi }+24gB_{2}\left( \theta \right) \right] =0.  \label{vs6}
\end{equation}%
The roots of Eq.~(\ref{vs6}) represent possible vacuum states and are given by%
\begin{equation}
\Psi =0,\qquad\qquad\Psi _{\pm }=\pm \sqrt{-\frac{1}{4\lambda _{\psi }L^{2}}%
\left[ \lambda _{\psi }+24gB_{2}\left( \theta \right) \right] }.  \label{vs7}
\end{equation}%
In order to know which solution above may be a physical vacuum state, one has to analyze
the stability of the effective potential \eqref{vs5}. This is achieved by means of its second derivative, i.e.,%
\begin{equation}
\frac{d^{2}V^{R}_{\mathrm{eff}}\left( \Psi \right) }{d\Psi ^{2}}=\frac{\lambda _{\psi }}{2}\Psi ^{2}+\frac{1}{24L^{2}}\left[ \lambda
_{\psi }+24gB_{2}\left( \theta \right) \right] .  \label{vs8}
\end{equation}%
For the vacuum state to be stable the second derivative of the potential,
evaluated at \eqref{vs7}, must be greater than zero. In this sense, we
investigate for which values of the parameter $\theta$ of the quasiperiodic condition and
of the coupling constants the stability is achieved.

\begin{figure}[h]
\includegraphics[scale=0.4]{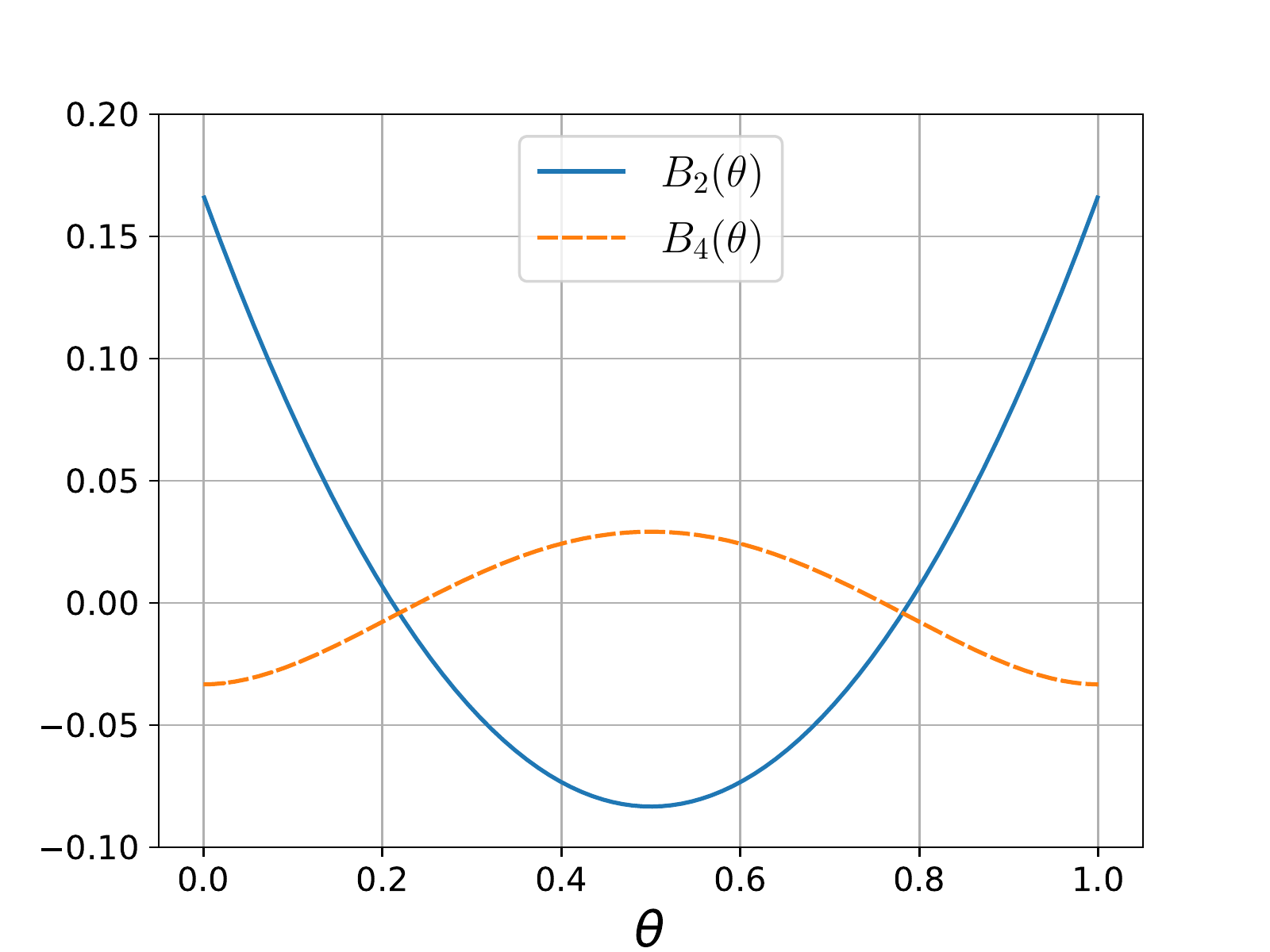}
\caption{Bernoulli polynomials $B_2(\theta)$, solid line, and $B_4(\theta)$, dashed line, as functions of the quasiperiodic parameter $\theta$.}\label{Bern}
\end{figure}

Let us then first consider the vacuum state, $\Psi =0$, which is the
case considered previously in the analysis of the Casimir energy density, its loop 
correction and the topological mass. Hence, from Eq.~(\ref{vs8}), one sees that $\Psi =0$ is
stable only if the following condition is satisfied:
\begin{equation}
\lambda _{\psi }>-24gB_{2}\left( \theta \right) .  \label{vs9}
\end{equation}%
As we can see, the parameter $\theta $ plays a crucial role in determining whether or not this vacuum
state is stable. Additionally, the coupling constants has also a great influence in the
vacuum stability. Thus, by taking the coupling constants, $\lambda _{\psi }$
and $g$, to be positive, and for $B_{2}\left( \theta \right)$ also positive, the condition above is always satisfied. However, if
$B_{2}\left( \theta \right)$ is negative the condition in Eq. \eqref{vs9} may be violated if $\lambda _{\psi }<g$. In Fig.\ref{Bern} we have
plotted the Bernoulli polynomial $B_{2}\left( \theta \right)$, from where we can see its positive and negative values.

By using the explicit form of the Bernoulli polynomial, that is, 
$B_{2}\left( \theta \right) =\theta ^{2}-\theta +\frac{1}{6}$, one finds that its negative values are provided  
for values of $\theta$ in the interval
\begin{equation}
\frac{1}{2}-\frac{\sqrt{3}}{6}<\theta <\frac{1}{2}+\frac{\sqrt{3}}{6}.
\label{vs10}
\end{equation}%
Of course, the values of $\theta$ for which $B_2(\theta)$ is positive reside out of the above interval. 

As an example, let us consider the particular case where $\theta =0.5$. Hence, the condition of
stability written in Eq.~(\ref{vs9}) becomes
\begin{equation}
\lambda _{\psi }>2g.  \label{vs11}
\end{equation}%
The above condition is in agreement with the one found in \cite%
{toms1980interacting}. Note, however, that we are considering a complex scalar field. 
The topological mass, in this case, takes the following form:
\begin{equation}
m_{T}^{2}=\frac{\left( \lambda _{\psi }-2g\right) }{24L^{2}},  \label{vs12}
\end{equation}%
which agrees with the result obtained in Eq.~(\ref{rcq6}) for $\theta =0.5$. 
Note also that if the interaction between the fields is not presente, that
is, $g=0$, we recover the previous known result found in Refs. \cite{Ford:1978ku, toms1980symmetry}.

According to Eq.~(\ref{vs7}) it is also possible to consider, $\Psi =\Psi_{\pm }$,
as the vacuum states, instead of $\Psi =0$. In this case,
evaluating the second derivative of the potential in Eq.~(\ref{vs8}), at $\Psi
=\Psi _{\pm }$, and setting the result to be greater than zero, one obtains
the vacuum stability condition as
\begin{equation}
\lambda _{\psi }<-24gB_{2}\left( \theta \right) .  \label{vs13}
\end{equation}%
For positive coupling constants, the above condition is satisfied only if
the Bernoulli polynomial, $B_{2}\left( \theta \right)$, is negative. This
is in fact possible for values of $\theta$ in the interval \eqref{vs10}. By considering the same
example as before, that is, $\theta =0.5$, we obtain the stability condition for a twisted scalar field 
as
\begin{equation}
\lambda _{\psi }<2g.  \label{vs14}
\end{equation}%
Consequently, the topological mass for this case reads
\begin{equation}
m_{T}^{2}=\frac{2g-\lambda _{\psi }}{12L^{2}}.  \label{vs15}
\end{equation}%
Note that the above result for the topological mass differs from the one
presented in Eq.~(\ref{rcq6}) for $\theta =0.5$. This difference arises from
the fact that the considered stable vacuum state for Eq.~(\ref{vs15}), that is, $%
\Psi =\Psi _{\pm }$, is not the same as the one in Eq.~(\ref{rcq6}), that is, $%
\Psi =0$. For the latter to be stable, as we have seen above, it is necessary
to consider the restriction in Eq. \eqref{vs9}. The result in Eq. \eqref{vs15}
is in agreement with the one found in Ref. \cite{toms1980symmetry}.

The Casimir energy density can also be obtained by taking, $\Psi =\Psi _{\pm }$,
as the stable vacuum state. Thus, from Eq. (\ref{vs7}), the effective potential given by Eq~(%
\ref{vs5}) provides the following expression for the Casimir energy
density:
\begin{eqnarray}
\mathcal{E}_C&=&\left. V^{R}_{\mathrm{eff}}\left( \Psi \right) \right\vert
_{\Psi _{\pm }}\nonumber\\
&\simeq&-\frac{\pi ^{2}}{90L^{4}}+2\frac{\pi ^{2}}{3L^{4}}%
B_{4}\left( \theta \right) -\frac{1}{384\lambda _{\psi }L^{4}}\left[ \lambda
_{\psi }+24gB_{2}\left( \theta \right) \right] ^{2}.  \label{vs16}
\end{eqnarray}%
Note that the first two terms on the r.h.s. of Eq.~(\ref{vs16}) are in
agreement with the Casimir energy density presented in Eq.~(\ref{rc1.16}).
However, the third term presents a dependency on the coupling constants $%
\lambda _{\psi }$ and $g$, which does not appear in the case where the
stable vacuum state is $\Psi =0$. It is important to point out that Eq. \eqref{vs16}
is only an approximation, up to first order in the coupling 
constants, since we are taking into account the expansion of
the effective potential presented in Eq.~(\ref{vs5}). The first and third terms
are always negative, while the second term can be positive or negative, depending on
the value of the Bernoulli Polynomial $B_{4}\left( \theta \right) $, shown in Fig.\ref{Bern}.

In order to calculate the two-loop correction contribution to the 
Casimir energy density in Eq. \eqref{vs16} would be necessary to
consider additional Feynman graphs other than the ones shown
in Figs.\ref{figure4}, \ref{figure5}. These additional Feynman graphs come from the second
term on the r.h.s. of Eq. (24) in Ref. \cite{toms1980symmetry},
which vanishes in the case $\Psi=0$ is the stable vacuum state.
The consideration of the two-loop contribution, of course, would make our problem extremely difficult so that
we restrict our analysis only to the one-loop correction that provides 
the Casimir energy density in Eq. \eqref{vs16}.

From Eqs. (\ref{vs9}) and (\ref{vs13}), we can conclude that the
stability of the vacuum states is determined by the values of the coupling
constants $\lambda _{\psi }$ and $g$, as well as by the value of the
parameter $\theta $ of the quasiperiodic condition for the complex field. However
there is no dependency on the parameter $L$.

In the next section we consider the same system as the one considered in
this section, but the complex field is now subjected to mixed boundary
conditions.

\section{Periodic condition and mixed boundary conditions}
%
\label{sec3.2}In this section we consider the real scalar field obeying periodic condition
as before, but now the complex scalar field is subject to mixed boundary conditions. In practice, the first order loop correction to the
effective potential associated with the real scalar field is the same as in Eq.~(\ref{rc1.3}), 
differently from the complex field which yields a different contribution since it obeys a
different condition. In this case, it is sufficient to evaluate
only the correction associated with the complex field. We will also assume that $\Psi=0$
is the stable vacuum state for the analysis below, although a discussion of other possible stable vacuum states 
is given in Sec.\ref{VSMB}.

\begin{figure}[h]
\includegraphics[scale=0.25]{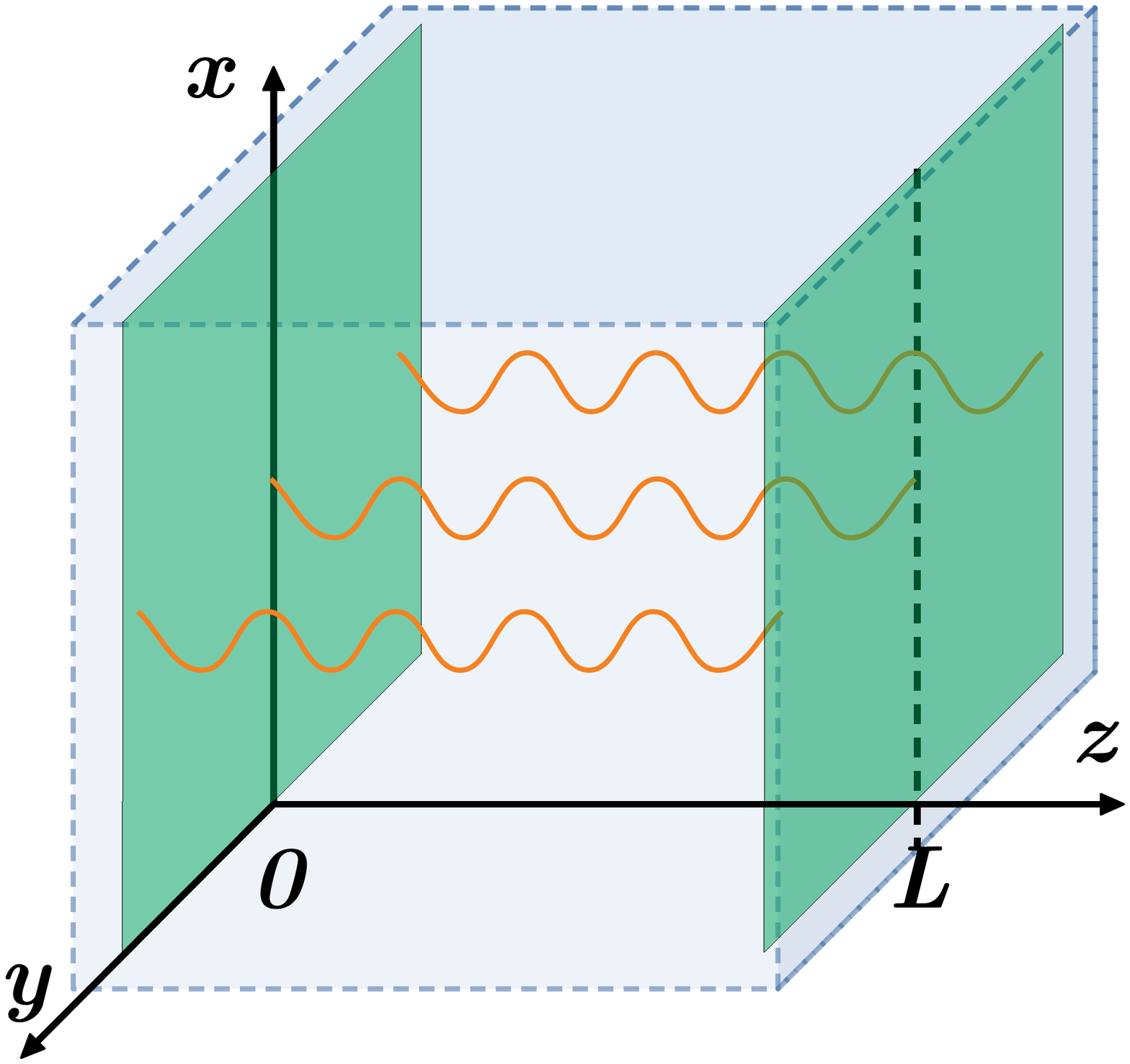}
\caption{Two identical and perfectly reflecting parallel planes placed at $z=0$ and $z=L$, confining the field modes of a complex scalar field. On the planes
the mixed boundary conditions in Eq. \eqref{m1} are applied.}\label{figure6}
\end{figure}

The complex field real components are subject to the following mixed boundary conditions
applied on the planes shown in Fig.\ref{figure6} \cite{barone2004radiative,cruz2020casimir, Ferreira:2023uxs}:
\begin{equation}
\left. \varphi _{i}\left(w\right) \right\vert _{z=0}=\left. \frac{\partial
\varphi _{i}\left(w\right) }{\partial z}\right\vert _{z=L},\qquad\qquad \left. \frac{%
\partial \varphi _{i}\left( w\right) }{\partial z}\right\vert _{z=0}=\left.
\varphi _{i}\left( w\right) \right\vert _{z=L},  \label{m1}
\end{equation}%
where $w=(\tau, x, y, z)$. By taking into account the boundary conditions above, the eigenvalues of the operator $\hat{B}$
given in Eq.~(\ref{rc2.5}), takes the form \cite{barone2004radiative, cruz2020casimir}
\begin{equation}
\beta _{\rho }=k_{\tau }^{2}+k_{x}^{2}+k_{y}^{2}+\left( n+\frac{1}{2}%
\right) ^{2}\frac{\pi ^{2}}{L^{2}}+M_{g}^{2},\qquad\qquad M_{g}^{2}=\mu ^{2}+g\Psi
^{2}, \label{m2}
\end{equation}%
where $n=0,1,2,..., $ and the subscript $\rho $ stands for the set of quantum numbers $\left(
k_{\tau },k_{x},k_{y},n\right) $. It is worth pointing out that, from Eq. \eqref{m1}, two configurations
are possible on the parallel planes in Fig.\ref{figure6}. For the plane at $z=0$ we can have Dirichlet boundary 
condition while for the plane at $z=L$ we can have the Neumann one. Conversely, for the plane at $z=0$ we can have Neumann 
boundary condition while for the plane at $z=L$ we can have the Dirichlet one. However, both configurations
provide the same eigenvalues in Eq. \eqref{m2}.

Having the eigenvalues obtained in Eq. \eqref{m2} we can now proceed to the investigation of the first order correction, that is,
the one-loop correction to the effective potential associated with the complex scalar field subjected to mixed boundary conditions on the planes shown in Fig.\ref{figure6}.

\subsection{One-loop correction}
%
The required steps for the obtention of the generalized zeta function for
the case under consideration, goes in a similar way as the one presentend in
the previous sections and also in \cite{cruz2020casimir}. Therefore, we
present only the main steps for the reader's convenience. Constructing the
generalized zeta function with the eigenvalues presented in \eqref{m2} requires the use of the
identity in Eq.~(\ref{i1}) which, after the integration of the momenta, one
can use the integral representation of the gamma function, (\ref{intg}),
finding the expression%
\begin{equation}
\zeta _{\beta }\left( s\right) =\frac{\Omega _{4}\pi ^{\frac{3}{2}-2s}}{%
8L^{4-2s}}\frac{\Gamma \left( s-\frac{3}{2}\right) }{\Gamma \left( s\right) }%
\sum_{n=0}^{+\infty }\left[ \left( n+\frac{1}{2}\right) ^{2}+\left( \frac{%
M_{g}L}{\pi }\right) ^{2}\right] ^{\frac{3}{2}-s},  \label{m3}
\end{equation}%
where $\Omega _{4}$ is the 4-dimensional volume written as $\Omega _{4}=\Omega _{3}L$, with $\Omega _{3}$ being the 3-dimensional volume associated with the Euclidean spacetime coordinates $\tau,x,y$. In order to perform the sum in Eq.~(\ref{m3}), we write it as a sum of two
terms \cite{cruz2020casimir, bordag2009advances}, i.e.,%
\begin{equation}
\sum_{n=0}^{+\infty }\left[ \left( n+\frac{1}{2}\right) ^{2}+\vartheta ^{2}%
\right] ^{\frac{3}{2}-s}=\frac{1}{2^{3-2s}}\left\{ \sum_{n=1}^{\infty }\left[
n^{2}+\left( 2\vartheta \right) ^{2}\right] ^{\frac{3}{2}-s}-2^{3-2s}%
\sum_{n=1}^{\infty }\left[ n^{2}+\vartheta ^{2}\right] ^{\frac{3}{2}%
-s}\right\} .  \label{m4}
\end{equation}%
Each sum on the r.h.s. of Eq.~(\ref{m4}) can be written in terms of the
Epstein-Hurwitz zeta function \cite{gradshteyn2014table}%
\begin{eqnarray}
\zeta _{EH}\left( z,\kappa \right) &=&\sum_{n=1}^{+\infty }\left( n^{2}+\kappa
^{2}\right) ^{-z}\nonumber\\
&=&-\frac{\kappa ^{-2z}}{2}+\frac{\pi ^{\frac{1}{2}}}{2}\frac{%
\Gamma \left( z-\frac{1}{2}\right) }{\Gamma \left( z\right) }\kappa ^{1-2z}+%
\frac{2^{1-z}\left( 2\pi \right) ^{2z-\frac{1}{2}}}{\Gamma \left( z\right) }%
\sum_{n=1}^{\infty }n^{2z-1}f_{\left( z-\frac{1}{2}\right) }\left( 2\pi
n\kappa \right) .  \label{m5}
\end{eqnarray}%
Hence, with the help of the Eq.~(\ref{m5}) one obtains the generalized zeta
function as%
\begin{equation}
\zeta _{\beta }\left( s\right) =\frac{\Omega _{4}}{16\pi ^{2}\Gamma \left(
s\right) }\left\{ M_{g}^{4-2s}\Gamma \left( s-2\right) +\frac{2^{s}}{L^{4-2s}%
}\sum_{n=1}^{\infty }n^{2s-4}\left[ 2^{2s-3}f_{\left( s-2\right) }\left(
4nM_{g}L\right) -f_{\left( s-2\right) }\left( 2nM_{g}L\right) \right]
\right\} .  \label{m6}
\end{equation}%
Evaluating the above expression and its derivative in the limit $%
s\rightarrow 0$, one finds the complex field contribution to the first order loop correction to the effective potential from
Eq.~(\ref{rc14}), i.e.,%
\begin{equation}
V_{\beta }^{\left( 1\right) }\left( \Psi \right) =\frac{M_{g}^{4}}{32\pi
^{2}}\left[ \ln\left( \frac{M_{g}^{2}}{\nu ^{2}}\right)-\frac{3}{2}\right] -\frac{%
M_{g}^{4}}{\pi ^{2}}\sum_{n=1}^{\infty }\left[ 2f_{2}\left( 4nM_{g}L\right)
-f_{2}\left( 2nM_{g}L\right) \right] .  \label{m7}
\end{equation}

Taking into consideration the contribution from the real field, Eq.~(\ref{rc1.3}), 
along with the contribution above of the complex field,
the effective potential, up to one-loop correction, is presented in the form%
\begin{eqnarray}
V_{\mathrm{eff}}\left( \Psi \right) &=&\frac{m^{2}+C_{2}}{2}\Psi ^{2}+\frac{%
\lambda _{\psi }+C_{1}}{4!}\Psi ^{4}+C_{3}+  \notag \\
&&+\frac{M_{\lambda }^{4}}{64\pi ^{2}}\left[ \ln \left( \frac{M_{\lambda
}^{2}}{\nu ^{2}}\right) -\frac{3}{2}\right] +\frac{M_{g}^{4}}{32\pi ^{2}}%
\left[ \ln\left( \frac{M_{g}^{2}}{\nu ^{2}}\right)-\frac{3}{2}\right] +  \notag \\
&&-\frac{M_{\lambda }^{4}}{2\pi ^{2}}\sum_{j=1}^{\infty }f_{2}\left(
jM_{\lambda }L\right) -\frac{M_{g}^{4}}{\pi ^{2}}\sum_{n=1}^{\infty }\left[
2f_{2}\left( 4nM_{g}L\right) -f_{2}\left( 2nM_{g}L\right) \right] .
\label{m8}
\end{eqnarray}%
Knowing the effective potential expressed in Eq.~(\ref{m8}), one needs to
renormalize it. Hence, by applying the renormalization conditions given by 
Eqs.~(\ref{rc14.2}), (\ref{rc14.3}) and (\ref{rc14.5}), we find the
renormalization constants $C_{i}$ as the same as the ones obtained in Eq. \eqref{rcq3},
as it should be.
After the substitution of these constants $C_{i}$'s in the
effective potential, (\ref{m8}), one can write the renormalized effective
potential as%
\begin{eqnarray}
V_{\mathrm{eff}}^{R}\left( \Psi \right) &=&\frac{m^{2}}{2}\Psi ^{2}+\frac{%
\lambda _{\psi }}{4!}\Psi ^{4}+\frac{\mu ^{4}}{32\pi ^{2}}\ln\left( \frac{M_{g}^{2}%
}{\mu ^{2}}\right)+\frac{m^{4}}{64\pi ^{2}}\ln \left( \frac{M_{\lambda }^{2}}{m^{2}}%
\right) +  \notag \\
&&+\frac{g\mu ^{2}\Psi ^{2}}{16\pi ^{2}}\left[ \ln\left( \frac{M_{g}^{2}}{\mu ^{2}}\right)%
-\frac{1}{2}\right] +\frac{\lambda _{\psi }^{2}\Psi ^{4}}{256\pi ^{2}}\left[
\ln \left( \frac{M_{\lambda }^{2}}{m^{2}}\right) -\frac{3}{2}\right] + 
\notag \\
&&+\frac{\lambda _{\psi }m^{2}\Psi ^{2}}{64\pi ^{2}}\left[ \ln \left( \frac{%
M_{\lambda }^{2}}{m^{2}}\right) -\frac{1}{2}\right] +\frac{g^{2}\Psi ^{4}}{%
32\pi ^{2}}\left[ \ln\left( \frac{M_{g}^{2}}{\mu ^{2}}\right)-\frac{3}{2}\right] +  \notag
\\
&&-\frac{M_{\lambda }^{4}}{2\pi ^{2}}\sum_{j=1}^{\infty }f_{2}\left(
jM_{\lambda }L\right) -\frac{M_{g}^{4}}{\pi ^{2}}\sum_{n=1}^{\infty }\left[
2f_{2}\left( 4nM_{g}L\right) -f_{2}\left( 2nM_{g}L\right) \right] .
\label{m10}
\end{eqnarray}

Once we obtain the renormalized effective potential found in Eq.~(\ref{m10}), the
Casimir energy density is written in a straightforwardly way by setting $%
\Psi =0$, i.e.,%
\begin{equation}
\mathcal{E}_{\mathrm{C}}=\left. V_{\mathrm{eff}}^{R}\left( \Psi \right)
\right\vert _{\Psi =0}=-\frac{m^{4}}{2\pi ^{2}}\sum_{j=1}^{\infty
}f_{2}\left( jmL\right) -\frac{\mu ^{4}}{\pi ^{2}}\sum_{n=1}^{\infty }\left[
2f_{2}\left( 4n\mu L\right) -f_{2}\left( 2n\mu L\right) \right] .
\label{m11}
\end{equation}%
The first term on the r.h.s. of Eq.~(\ref{m11}) is the contribution from the
real field which is equal to the one in Eq.~(\ref{rc1.15}) as it should,
since the boundary condition applied to the real field is the same.
However, the second term on the r.h.s. of Eq.~(\ref{m11}) is the
contribution from the complex field and differs from the case of
quasiperiodic condition presented in Eq.~(\ref{rc1.15}). This contribution
is consistent with the result shown in Ref. \cite{barone2004radiative}
where the authors considered a self-interacting real scalar field. 

\begin{figure}[h]
\includegraphics[scale=0.3]{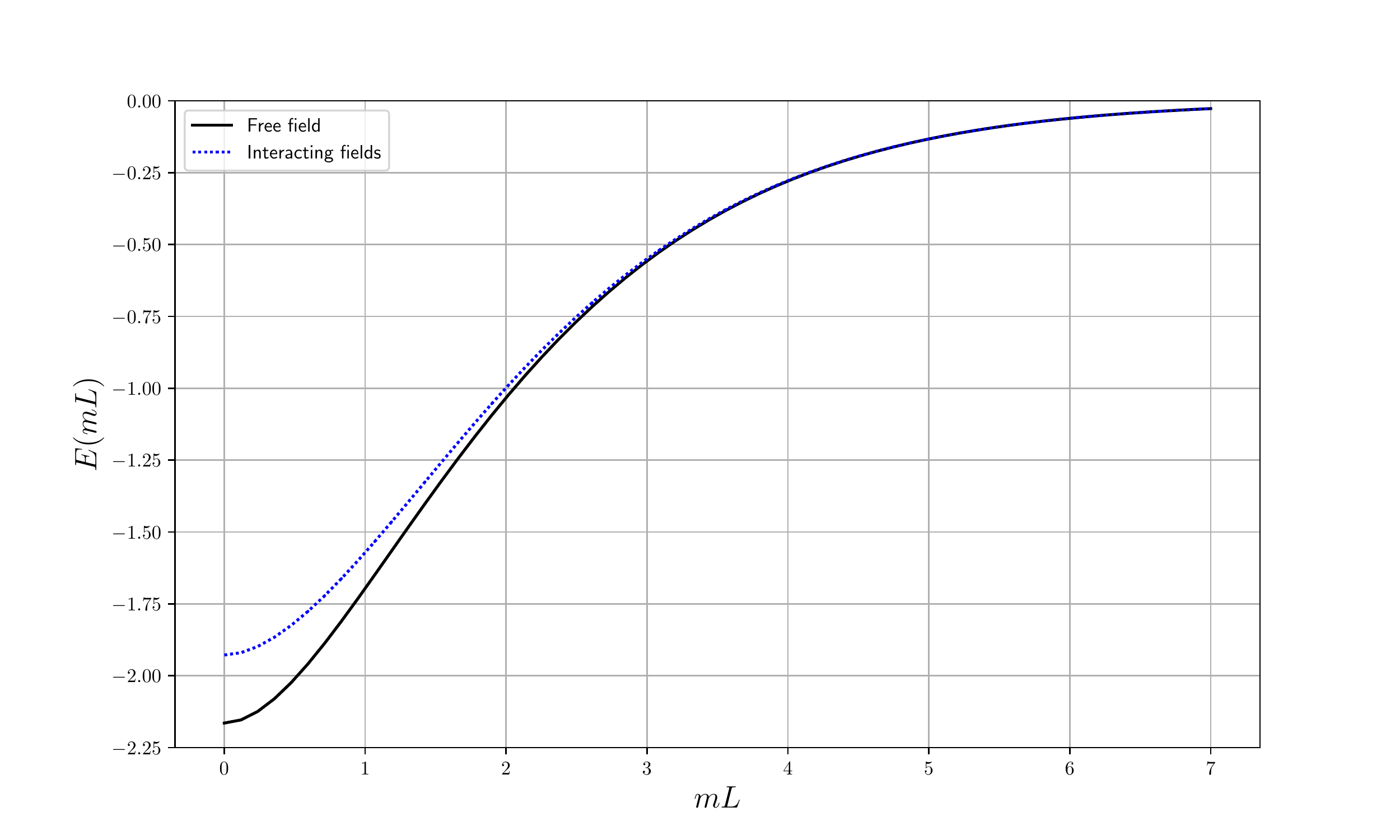}
\includegraphics[scale=0.3]{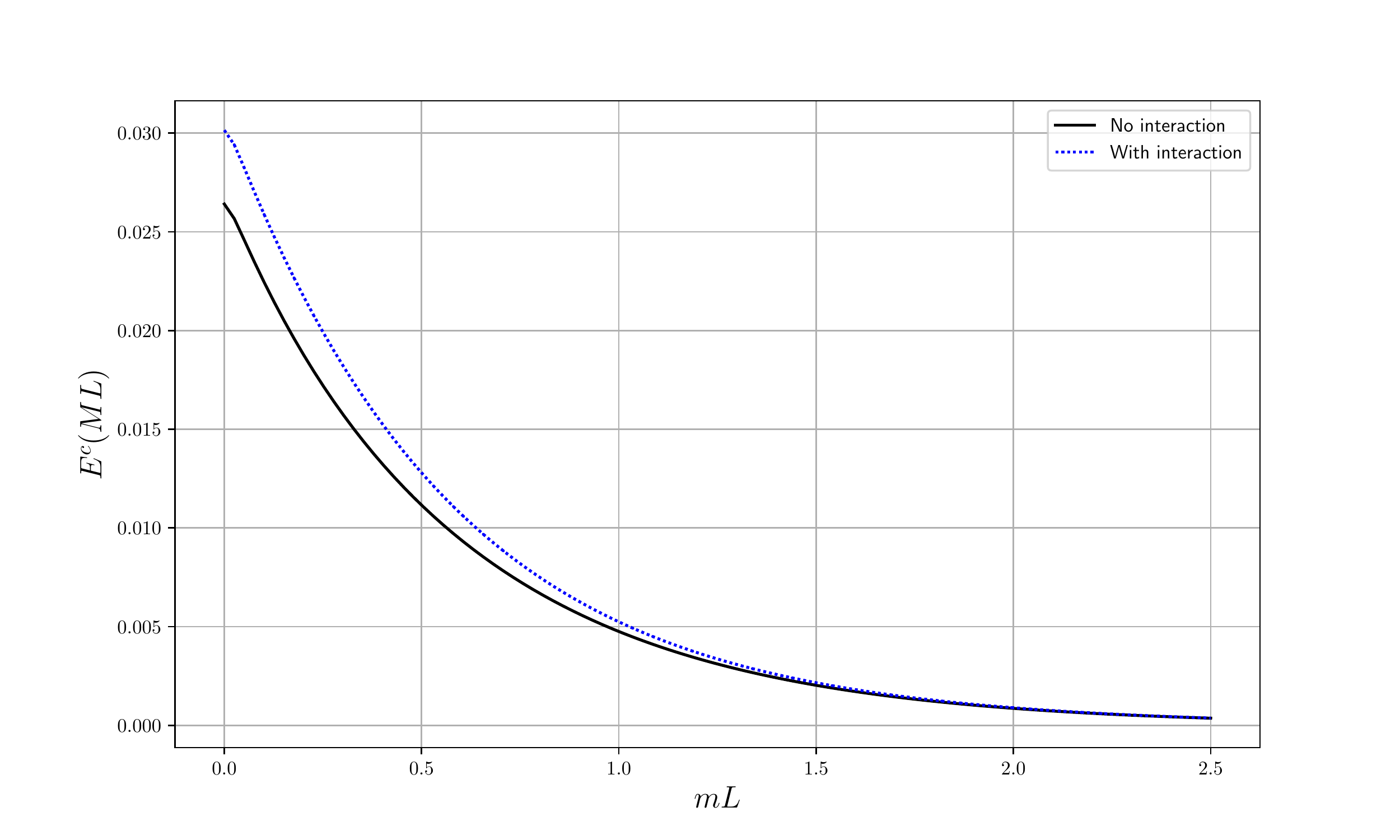}
\caption{Plot of the dimensionless Casimir energy density, $E(mL)=2\pi^2L^4\mathcal{E}_{\mathrm{C}}$, defined from Eq. (\ref{m11}), as a function of $mL$ is shown on the left, while the plot on the right shows the dimensionless two-loop contribution to the Casimir energy density, $E^c(mL)=32\pi^4L^4\Delta\mathcal{E}_{\mathrm{C}}$, defined from Eq. \eqref{msv17}, as a function of $mL$ and considering $\lambda_{\psi}=10^{-2}$, $\lambda_{\varphi}=10^{-2}$ and $g=10^{-3}$. For both cases we have taken $\mu=m$.}\label{figure7}
\end{figure}

The massless scalar field case
is obtained by taking the limit for small arguments of the Macdonald function \cite{abramowitz1965handbook}.
This yields the following Casimir energy density:%
\begin{eqnarray}
\mathcal{E}_{\mathrm{C}}&=&-\frac{\pi ^{2}}{90L^{4}}\left[ 1-\frac{7}{64}%
\right]\nonumber\\
&=&\frac{57}{64}\times\left(-\frac{\pi ^{2}}{90L^{4}}\right),
  \label{m12}
\end{eqnarray}
where we can see that the effect of the interaction with the complex field subjected to mixed boundary conditions is to increase the Casimir energy density of the 
real scalar field under a periodic condition.

In Fig.\ref{figure7} we have plotted the Casimir energy density in Eq. \eqref{m11} as a function of $mL$ and taken $m=\mu$, showing it on the left side. The latter shows how the curve for the free real scalar field (black solid line) differs from the curve when considering the influence of the interaction (blue dotted line). In fact, the interaction increases the value of the Casimir energy density, as shown the curves. This plot also shows that the Casimir energy density goes to zero for large values of $mL$. This a consequence of the exponentially suppressed behavior of the Macdonald function for large arguments \cite{abramowitz1965handbook}. Also, the two curves end in their corresponding massless field constant value cases at $mL=0$, as it can be checked from Eq. \eqref{m12}. 

Let us now investigate how the topological mass associated with the real field changes under the influence
of mixed boundary conditions imposed on the complex field. Thus, applying the renormalization condition (%
\ref{rc14.3}), with the renormalized effective potential in Eq.~(\ref{m10}),
provides the topological mass written as%
\begin{equation}
m_{T}^{2}=m^{2}\left\{ 1+\frac{\lambda _{\psi }}{4\pi ^{2}}%
\sum_{j=1}^{\infty }f_{1}\left( jmL\right) +\frac{\mu ^{2}}{m^{2}}\frac{g}{%
\pi ^{2}}\sum_{n=1}^{\infty }\left[ 2f_{1}\left( 4n\mu L\right) -f_{1}\left(
2n\mu L\right) \right] \right\} .  \label{m13}
\end{equation}%
Of course the difference between the above result and the topological mass
found in Eq.~(\ref{rcq5}), relies in the third term on the r.h.s. of Eq. \eqref{m13}.
Furthermore, by considering the massless scalar fields case, $m,\mu \rightarrow 0$, one
obtains the topological mass as follows%
\begin{equation}
m_{T}^{2}=\frac{2\lambda _{\psi }-g}{48L^{2}}.  \label{m14}
\end{equation}%
Note that the topological mass above coincides with the
particular cases of quasiperiodic condition in Eq.~(\ref{rcq6}), for $\theta
=1/4,3/4$, which are in the range of Eq.~(\ref{vs10}).

Similarly to the discussion presented in the previous section, here the topological mass squared can also become negative depending on whether $\lambda_{\psi}$ is bigger or smaller than $g$. For instance, if $2\lambda_{\psi}<g$, Eq. \eqref{m14} becomes negative, indicating vacuum instability. Again, had we considered a complex scalar field theory with only self-interaction (no interaction between the fields) this would be a problem since it does not make sense to consider a constant complex field, $\Phi_i\neq 0$, compatible with mixed boundary conditions. This problem is solved by taking into account an interaction theory as the one considered in the present section (see also \cite{toms1980interacting}). Within this theory it is possible to study the vacuum stability, which here is made in Sec.\ref{VSMB} for massless scalar fields. The analysis indicates that the vacuum $\Psi=0$ is stable only if $2\lambda_{\psi} > g$, otherwise it is necessary to consider the two other possible vacuum states, $\Psi_{\pm}$, in Eq. \eqref{mvs7}.

\begin{figure}[h]
\includegraphics[scale=0.3]{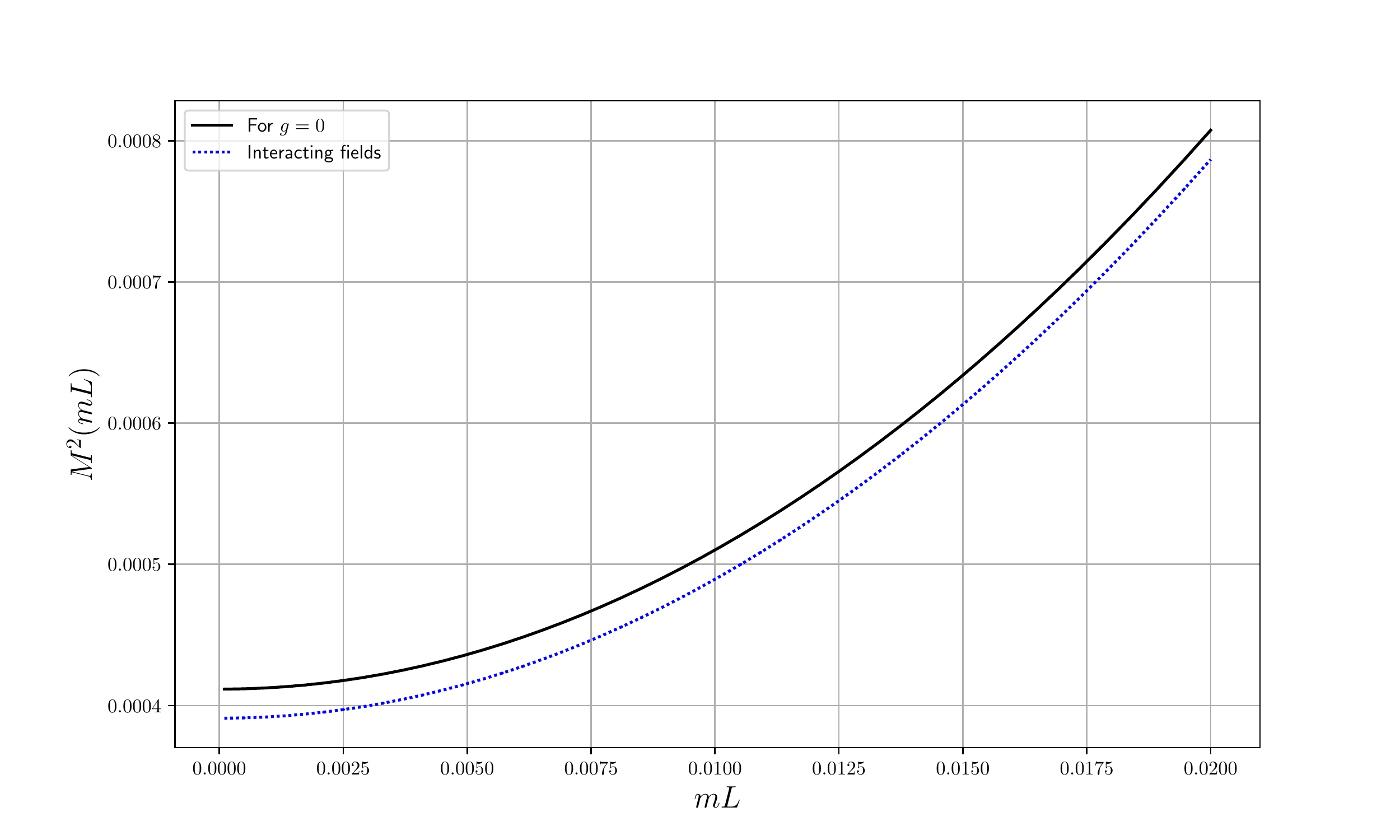}
\includegraphics[scale=0.3]{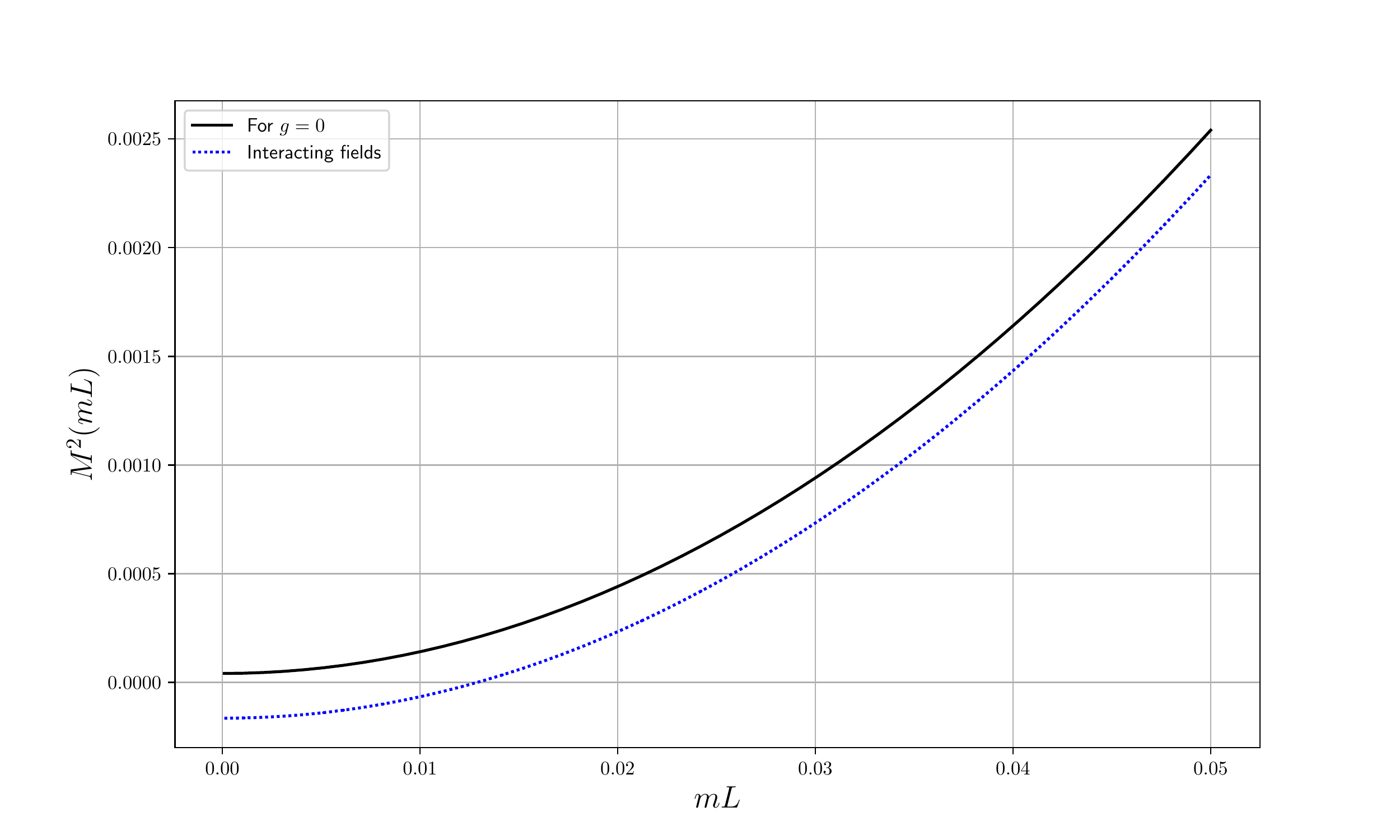}
\caption{Plot of the dimensionless topological mass squared, $M^2(mL)=m_T^2L^2$, defined from Eq. \eqref{m13}, as a function of $mL$, taken $\mu=m$. On the left, the plot shows the curves for $\lambda_{\psi}= 10^{-2}$ and $g= 10^{-3}$ while on the right the plot shows the curves for $\lambda_{\psi}= 10^{-3}$ and $g= 10^{-2}$.}\label{figure8}
\end{figure}

In Fig.\ref{figure8}, we have plotted the dimensionless mass squared, $M^2(mL) = m_T^2L^2$ , defined from Eq. \eqref{m13}, as a function of
$mL$, taken $\mu=m$. On the left of Fig.\ref{figure8} the plot shows the curves for $\lambda_{\psi} = 10^{-2}$ and $g = 10^{-3}$, which satisfies the condition $2\lambda_{\psi} > g$ in order $\Psi=0$ be a stable vacuum. In contrast, the plot on the right shows the curves for $\lambda_{\psi} = 10^{-3}$ and $g = 10^{-2}$, which satisfies the condition $2\lambda_{\psi} < g$ . In this case, Eq. \eqref{m14} becomes negative, showing that $\Psi=0$ is in fact an unstable vacuum state. Note that in an interacting massive scalar field theory
$\Psi=0$ may still be a stable vacuum state even if $2\lambda_{\psi} < g$ for large values of $mL$, as shown in the plot on the right side. Note also that each curve, at $mL=0$, end in their corresponding constant massless scalar field values for the topological mass in Eq. \eqref{m14}. For large values of $mL$ the Macdonald function is exponentially suppressed and the curves are dominated by the first term on the r.h.s. of Eq. \eqref{m13}. %

The one-loop correction analysis is now done, so one can proceed to the two-loop correction contribution by still considering $\Psi = 0$ as the stable vacuum state. As we now know, it means that we have to consider the restriction $2\lambda_{\psi} > g$. The vacuum stability analysis for the present case we postpone until Sec.\ref{VSMB}. 

\subsection{Two-loop correction}
Now we wish to evaluate the two-loop correction to the
effective potential. We use the same graphs as the ones used in the case of
quasiperiodic condition in Figs.\ref{figure4} and \ref{figure5}, and also a similar notation as the one
used in Sec.\ref{sec3.3}. The first correction comes from
the self-interaction term of the real scalar field, that is, $\frac{\lambda _{\psi }%
}{4!}\psi ^{4}$. Since we are interested in the vacuum state, $\Psi=0$, the only
non-vanishing contribution is the same as the one obtained in Eq.~(\ref%
{rcq9}).

The second contribution comes from the self-interaction of the complex
scalar field, i.e., $\frac{\lambda _{\varphi }}{4!}\varphi _{i}^{4}$. For the case
under consideration, this contribution reads,
\begin{equation}
V_{\lambda _{\varphi }}^{\left( 2\right) }\left( 0\right) =2\frac{\lambda
_{\varphi }}{8}\left. \left[ \frac{\zeta _{\gamma }^{R}\left( 1\right) }{%
\Omega _{4}}\right] ^{2}\right\vert _{\Psi =0}=2\frac{\lambda _{\varphi }\mu
^{4}}{32\pi ^{4}}\left\{ \sum_{n=1}^{\infty }\left[ 2f_{1}\left( 4n\mu
L\right) -f_{1}\left( 2n\mu L\right) \right] \right\} ^{2}.  \label{msv14}
\end{equation}%
Note that in the above expression, $\zeta _{\beta }^{R}\left( 1\right) $, stands
for the non-divergent part of $\zeta _{\beta }\left( s\right) $ given by
Eq.~(\ref{m6}), at $s=1$, and the factor of two in front of the constant $\lambda
_{\varphi }$ is to remind ourselves that we are taking into account the two
components of the complex field.

Next we obtain the contribution from the interaction between the fields,
that is, from the term $\frac{g}{2}\varphi _{i}^{2}\psi ^{2}$. Hence, this term
yields the following correction to the effective potential:%
\begin{equation}
V_{g}^{\left( 2\right) }\left( 0\right) =2\frac{gm^{2}\mu ^{2}}{8\pi ^{4}}%
\left[ \sum_{j=1}^{\infty }f_{1}\left( jmL\right) \right] \left\{
\sum_{n=1}^{\infty }\left[ 2f_{1}\left( 4n\mu L\right) -f_{1}\left( 2n\mu
L\right) \right] \right\} .  \label{msv15}
\end{equation}%
The last correction to the effective potential comes from the interaction between 
the real components of the complex field (also a self-interaction), that is, from the
term $\frac{\lambda _{\varphi }}{4!}2\varphi _{1}^{2}\varphi _{2}^{2}$. Thus, one
is able to write it as%
\begin{equation}
V_{2\lambda _{\varphi }}^{\left( 2\right) }\left( 0\right) =\frac{\lambda
_{\varphi }\mu ^{4}}{48\pi ^{4}}\left\{ \sum_{n=1}^{\infty }\left[
2f_{1}\left( 4n\mu L\right) -f_{1}\left( 2n\mu L\right) \right] \right\}
^{2}.  \label{msv16}
\end{equation}

Therefore, from the results obtained in Eqs.~(\ref{rcq9}), (\ref{msv14}), (\ref%
{msv15}) and (\ref{msv16}), we may write the Casimir energy density, up to
second order correction, that is, up to two-loop correction, as follows%
\begin{eqnarray}
\Delta\mathcal{E}_{\mathrm{C}} &=&\left. V^{\left( 2\right) }\left( \Psi \right) \right\vert _{\Psi =0} \nonumber\\
&=&\frac{\lambda _{\psi }m^{4}}{32\pi ^{4}}\left[
\sum_{j=1}^{\infty }f_{1}\left( jmL\right) \right] ^{2}+\frac{\lambda
_{\varphi }\mu ^{4}}{12\pi ^{4}}\left\{ \sum_{n=1}^{\infty }\left[
2f_{1}\left( 4n\mu L\right) -f_{1}\left( 2n\mu L\right) \right] \right\}
^{2}+  \notag \\
&&+\frac{gm^{2}\mu ^{2}}{4\pi ^{4}}\left[ \sum_{j=1}^{\infty }f_{1}\left(
jmL\right) \right] \left\{ \sum_{n=1}^{\infty }\left[ 2f_{1}\left( 4n\mu
L\right) -f_{1}\left( 2n\mu L\right) \right] \right\} .  \label{msv17}
\end{eqnarray}%
Note that the expression above is proportional to the coupling constants $\lambda
_{\psi }$, $\lambda _{\varphi }$ and $g$, representing the self-interaction
of each field and also the interaction between the fields. Moreover, from
the correction to the Casimir energy density presented in Eq.~(\ref{msv17}), one can consider the
massless scalar fields limit, $m,\mu \rightarrow 0$. Recalling the limit of small
arguments for the Macdonald function, i.e., $K_{\mu }\left( x\right) \simeq 
\frac{\Gamma \left( \mu \right) }{2}\left( \frac{2}{x}\right) ^{\mu }$ \cite%
{abramowitz1965handbook}, one finds the correction in Eq. (\ref{msv17}) for the
massless fields in the form%
\begin{equation}
\Delta\mathcal{E}_{\mathrm{C}}=\frac{\lambda _{\psi }}{1152L^{4}}+\frac{\lambda _{\varphi }}{%
27648L^{4}}-\frac{g}{1152L^{4}}.  \label{msv18}
\end{equation}%
As we can see, the corrections proportional to the coupling constants $%
\lambda _{\psi }$ and $\lambda _{\varphi }$, which come from the
self-interaction of the fields, increase the Casimir energy density in Eq. (\ref%
{m12}) while the term coming from the interaction between the fields, codified by 
the coupling constant $g$, have the effect of
decrease the Casimir energy density. Note that the contribution proportional to $\lambda _{\varphi }$
present in Eq. \eqref{msv18} is not the same as the one obtained in Ref. \cite{barone2004radiative}
for the self-interacting real scalar field. In fact, our result for the second term on the r.h.s. of Eq.  \eqref{msv18} 
is 8/3 bigger than the one obtained in Ref. \cite{barone2004radiative}. This is due to the fact
that, besides the contribution in Eq. \eqref{msv14}, we also have an additional contribution
proportional to $\lambda _{\varphi }$ coming from the interaction between the components of 
the complex field in Eq. \eqref{msv16}. The same is valid for the massive contribution on the second
term on the r.h.s. of Eq. \eqref{msv17}. Note also that, in order to compare our results with
the ones present in Ref. \cite{barone2004radiative} we need to define the Casimir energy correction, $\Delta E_C$,
per unit area, $A$, of the planes as $\frac{\Delta E_C}{A}=L\Delta\mathcal{E}_{\mathrm{C}}$.

In Fig.\ref{figure7}, the plot on the right shows the influence of the complex field, under mixed boundary conditions, on the correction \eqref{msv17} to the Casimir energy density of a massive real scalar field. The expression in Eq. \eqref{msv17} has been plotted as a function of $mL$, taken $\mu=m$. We also have considered $\lambda_{\psi} = 10^{-2}$, $\lambda_{\varphi} = 10^{-2}$ and $g=10^{-3}$. The black solid line is the correction free of interaction with the complex field, only with the effect of the real field self-interaction, while the blue dotted line is the correction \eqref{msv17} taking into account the interaction with the complex field subjected to mixed boundary conditions. The effect of the latter is to increase the correction, as revealed by the plot in Fig.\ref{figure7}. Note that the two curves tend to their corresponding massless field constant value cases at $mL=0$, as it can be checked from Eq. \eqref{msv18}. Also, in the regime, $mL\gg1$, the correction in Eq. \eqref{msv17} goes to zero. This is once again a consequence of the exponentially suppressed behavior of the Macdonald function for large arguments \cite{abramowitz1965handbook}.

Next, we shall analyze the vacuum stability of the theory, since the state $\Psi=0$ is not the only possible vacuum state, as we have already mentioned. For simplicity, we shall consider a massless scalar field theory.

\subsection{Vacuum stability}
\label{VSMB}

Let us analyze here the stability of the possible vacuum states associated with the effective potential, up to first order loop correction, of the theory described by the action in Eq. \eqref{rc2}. For simplicity we consider the case where the fields are massless, i.e., $\mu, m\rightarrow 0$. It is import to point out again that, for the complex scalar field obeying the boundary conditions in Eq. \eqref{m1}, the only constant field that can satisfy such a condition is the zero field, hence, we set $\Phi_i=0$. As mentioned before, this fact also turns the approximation discussed below Eq. \eqref{dAl} into an exact expression, namely, the one in Eq. \eqref{rc2.3}, which does not consider cross terms.

By following the same steps as the ones to obtain Eq. \eqref{rcq2}, the nonrenormalized effective potential for the massless scalar fields case is written as
\begin{eqnarray}
&&V_{\mathrm{eff}}\left( \Psi \right) =\frac{\lambda _{\psi }+C}{4!}\Psi
^{4}+\frac{\lambda _{\psi }^{2}\Psi ^{4}}{256\pi ^{2}}\left[ \ln \left( 
\frac{\lambda _{\psi }\Psi ^{2}}{2\nu ^{2}}\right) -\frac{3}{2}\right] +%
\frac{g^{2}\Psi ^{4}}{32\pi ^{2}}\left[ \ln \left( \frac{g\Psi ^{2}}{\nu ^{2}%
}\right) -\frac{3}{2}\right] +  \notag \\
&&-\left( \frac{\lambda _{\psi }}{2}\right) ^{2}\frac{\Psi ^{4}}{2\pi ^{2}}%
\sum_{j=1}^{\infty }f_{2}\left( j\sqrt{\frac{\lambda _{\psi }}{2}\Psi ^{2}}%
L\right) -\frac{g^{2}\Psi ^{4}}{\pi ^{2}}\sum_{n=1}^{\infty }\left[
2f_{2}\left( 4n\sqrt{g\Psi ^{2}}L\right) -f_{2}\left( 2n\sqrt{g\Psi ^{2}}%
L\right) \right] .  \label{mvs1}
\end{eqnarray}%
Now it is required to renormalize the effective potential in Eq.~(\ref{mvs1}).
In this sense, by applying the renormalization condition given by Eq.~(\ref{rc14.2}),
one obtains the renormalization constant $C$ as the same as the one in Eq. \eqref{vs3}.
Thus, by substituting this renormalization constant into the
effective potential presented in Eq.~(\ref{mvs1}), yields the renormalized
effective potential, i.e.,%
\begin{eqnarray}
&&V_{\mathrm{eff}}^{R}\left( \Psi \right) =\frac{\lambda _{\psi }}{4!}\Psi
^{4}-\left[ \frac{\lambda _{\psi }^{2}}{8}+g^{2}\right] \frac{25\Psi ^{4}}{%
192\pi ^{2}}+\left[ \frac{\lambda _{\psi }^{2}}{8}+g^{2}\right] \frac{\Psi
^{4}}{32\pi ^{2}}\ln \left( \frac{\Psi ^{2}}{M ^{2}}\right) +  \notag \\
&&-\left( \frac{\lambda _{\psi }}{2}\right) ^{2}\frac{\Psi ^{4}}{2\pi ^{2}}%
\sum_{j=1}^{\infty }f_{2}\left( j\sqrt{\frac{\lambda _{\psi }}{2}\Psi ^{2}}%
L\right) -\frac{g^{2}\Psi ^{4}}{\pi ^{2}}\sum_{n=1}^{\infty }\left[
2f_{2}\left( 4n\sqrt{g\Psi ^{2}}L\right) -f_{2}\left( 2n\sqrt{g\Psi ^{2}}%
L\right) \right] .  \label{mvs4}
\end{eqnarray}%
In order to analyze the vacuum stability the renormalized effective potential, (\ref{mvs4}), can be expanded in terms of
the coupling constant $\lambda _{\psi },g$, keeping the terms only to first
order. This results in the following expression:%
\begin{equation}
V_{\mathrm{eff}}^{R}\left( \Psi \right) \simeq-\frac{19\pi ^{2}%
}{30L^{4}} +\frac{\lambda _{\psi }\Psi ^{4}}{4!}+%
\frac{\lambda _{\psi }\Psi ^{2}}{48L^{2}}-\frac{g\Psi ^{2}}{96L^{2}}.
\label{mvs6}
\end{equation}

The possible vacuum states are obtained as the value of $\Psi $ which
corresponds to the minimum of the expanded effective potential in Eq.~(\ref%
{mvs6}). Therefore, by deriving the effective potential in Eq.~(\ref{mvs6}) with respect
to $\Psi $ and equating it to zero, gives the following values of $\Psi $,
which correspond to the possible vacuum states:%
\begin{equation}
\Psi =0,\qquad\qquad \Psi _{\pm }=\pm \sqrt{\frac{g-2\lambda _{\psi }}{8\lambda
_{\psi }L^{2}}}.  \label{mvs7}
\end{equation}%
Whether the vacuum states presented in Eq.~(\ref{mvs7}) are stable or not,
is decided from the second derivative of the expandend effective potential
given in Eq.~(\ref{mvs6}).

Let us first consider the vacuum state as $\Psi =0$. Then, by taking
the second derivative of the expanded potential in Eq. (\ref{mvs6}), evaluated
at $\Psi=0$, one finds that the condition for the vacuum stability is presented as%
\begin{equation}
2\lambda _{\psi }>g.  \label{mvs9}
\end{equation}%
For this vacuum state the Casimir energy density is given by the same
expression as the one in Eq.~(\ref{m12}). Moreover, it is straightforward to see that the
topological mass also does not change, that is, it gives the same result as
the one in Eq.~(\ref{m14}).

From Eq. (\ref{mvs7}), on the other hand, one can also consider the vacuum state as $%
\Psi =\Psi _{\pm }$. By evaluating the second derivative of the expanded effective
potential at the vacuum states $\Psi =\Psi _{\pm }$, we learn that the stability
condition reads%
\begin{equation}
g>2\lambda _{\psi }.  \label{mvs10}
\end{equation}%
Hence, the topological mass for the case under consideration takes the following
form:%
\begin{equation}
m_{T}^{2}=\frac{g-2\lambda _{\psi }}{24L^{2}},  \label{msv11}
\end{equation}%
which is also a consistent quantity since it is strictly positive, in accordance
with Eq.~(\ref{mvs10}). Besides, the Casimir energy density, considering $%
\Psi =\Psi _{\pm }$ as the vacuum states, is obtained by the substitution of $%
\Psi _{\pm }$ on the expanded effective potential in Eq.~(\ref{mvs6}), providing the Casimir energy density%
\begin{eqnarray}
\mathcal{E}_C&=&\left. V^{R}_{\mathrm{eff}}\left( \Psi \right) \right\vert
_{\Psi _{\pm }}\nonumber\\
&\simeq&-\frac{19\pi ^{2}}{30L^{4}}-\frac{\left( g-2\lambda _{\psi }\right) ^{2}}{1536\lambda _{\psi
}L^{4}}.  \label{msv12}
\end{eqnarray}
We emphasize that the above result is an approximation since we are
considering the expansion of the effective potential in Eq.~(\ref{mvs6}) up 
to first order in the coupling constants. Note that the first term on the r.h.s. of
Eq. \eqref{msv12} is the Casimir energy density obtained in Eq. \eqref{m12}
while the second term brings coupling constant corrections.

The discussion presented at the end of Sec.\ref{sec3.1} also applies here. That is, the calculation of the two-loop correction contribution to the Casimir energy density in Eq. \eqref{msv12} requires additional Feynman graphs other than the ones shown in Figs.\ref{figure4}, \ref{figure5}. The consideration of the two-loop contribution, of course, would make our problem extremely difficult so that we restrict our analysis only to the one-loop correction that provides the Casimir energy density in Eq. \eqref{msv12}.

From the results presented in Eqs.~(\ref{mvs9}) and (\ref{mvs10}), one can
conclude that the stable vacuum state is determined by the values of the coupling
constants $\lambda _{\psi }$ and $g$ and not on the value of the parameter $L$
characterizing the boundary condition.

\section{Concluding remarks}
\label{sec4}
%
Loop correction to the Casimir effect and generation of topological mass have been
investigated. Both the Casimir energy density and the topological mass arise from the
nontrivial topology of the Minkowski spacetime, which takes place in the form of periodic and quasiperiodic
conditions. These physical quantities also arise from mixed boundary conditions considered. More specifically, the system that has been taken into consideration consists of real and complex scalar fields interacting by means of a quartic interaction in addition to the self-interactions of the fields.

The real scalar field has been subjected to a
periodic boundary condition while the complex scalar field has been assumed to satisfy
quasiperiodic condition, and also mixed boundary conditions.
The Casimir energy density, up to one-loop correction to the effective potential, has been obtained in Eq.~(\ref%
{rc1.15}) for massive fields, and in Eq.~(\ref{rc1.15.a}), for massless fields, considering the
case where the complex field obeys quasiperiodic condition. In
this context, the topological mass has also been obtained in Eqs.~(\ref{rcq5}) and (%
\ref{rcq6}) for the massive and massless cases, respectively. The two-loop
correction contribution to the Casimir energy density, considering both massive and massless
fields cases have been presented, respectively, in Eqs.~(\ref{c2l4}) and (\ref{c4ims}), which
turn out to be proportional to the coupling constants $\lambda _{\psi }$, $%
\lambda _{\varphi }$ and $g$. Moreover, it has also been investigated the possible
stable vacuum states and the stability conditions for such states. These
vacuum states have been presented in Eq.~(\ref{vs7}) and the corresponding
stability conditions expressed in Eqs.~(\ref{vs9}) and (\ref{vs13}),
which depend on the values of the coupling constants $\lambda _{\psi }$, $g$
and on the parameter $\theta $ of the quasiperiodic condition.

Furthermore, by assuming that the complex field satisfies mixed boundary
conditions, the Casimir energy density, up to one-loop correction to the effective potential, for both
massive and massless field cases have been presented in Eqs.~(\ref{m11}) and (\ref%
{m12}), respectively. The topological mass analysis for such a
system has been performed as well and the results are given by Eqs. (\ref{m13}) and 
(\ref{m14}) for massive and massless fields, respectively. In this case, the two-loop correction 
contribution to the Casimir energy density has also been presented in Eqs.~(%
\ref{msv17}) and (\ref{msv18}), for massive and massless cases, respectively. The
investigation of vacuum stability has determined the possible vacuum
states and the condition to achieve stability in each case. From the
results presented in Eqs.~(\ref{mvs9}) and (\ref{mvs10}), we can conclude
that the stable vacuum is determined by the values of the coupling constants 
$\lambda _{\psi }$ and $g$ and not on the value of the parameter, $L$, of the
boundary condition. 

Therefore, by extending the analysis performed in Ref. \cite{toms1980interacting} to the complex field 
and considering other conditions we have also generalized the results found in Refs. \cite{toms1980symmetry, Ford:1978ku, porfirio2021ground, barone2004radiative, cruz2020casimir} 
for a self-interacting real scalar field theory.

{\acknowledgments} A.J.D.F.J would like to thank the Brazilian agency Coordination for the Improvement of Higher Education Personnel (CAPES) 
for financial support. The author H.F.S.M. is partially supported by the
Brazilian agency National Council for Scientific and Technological Development (CNPq) under grant N$\textsuperscript{o}$ 311031/2020-0.


\end{document}